\documentclass{aastex}
\usepackage{diagbox}
\usepackage{array}
\usepackage{spr-astr-addons}
\usepackage{url}
\RequirePackage{color}

\begin{document}

\title{Multidimensional analysis of Fermi GBM\\ gamma-ray bursts}
 
\shorttitle{Multidimensional analysis of Fermi GBM GBRs}
\shortauthors{Horv\'ath, Hakkila et al.}

\author{I.~Horv\'ath\altaffilmark{1}}
\email{horvath.istvan@uni-nke.hu}
  \and \author{J.~Hakkila\altaffilmark{2}}
   \and \author{Z.~Bagoly\altaffilmark{3} } 
  \and \author{L.V.~T\'oth\altaffilmark{3} } 
  \and \author{I.I.~R\'acz\altaffilmark{1}  }
  \and \author{S.~Pint\'er\altaffilmark{1}  }
  \and \author{B.G.~T\'oth\altaffilmark{1}  }
\email{horvath.istvan@uni-nke.hu}
\altaffiltext{1}{National University of Public Service, Budapest, Hungary. \ \ \ \ corresponding author: horvath.istvan@uni-nke.hu}
\altaffiltext{2}{College of Charleston, Charleston, SC, USA.}
\altaffiltext{3}{E\"otv\"os  University, Budapest, Hungary.
    \email{horvath.istvan@uni-nke.hu}}

\begin{abstract}
The Fermi GBM catalog provides a large database with
many measured variables that can be used to explore and verify
gamma-ray burst classification results.
We have used Principal Component Analysis and statistical clustering techniques to
look for clustering in a sample of 801 gamma-ray bursts described by
sixteen classification variables. 
The analysis recovers what appears to be the Short class
and two long-duration classes that differ from one another
via peak flux, with negligible variations in 
fluence, duration and spectral hardness. Neither class has properties 
entirely consistent with the Intermediate GRB class.
Spectral hardness has been a critical Intermediate class property. 
Rather than providing spectral hardness, Fermi GBM provides a range of 
fitting variables for four different spectral models; it is not intuitive how these variables
can be used to support or disprove previous GRB classification results. 

\end{abstract}


\maketitle 

\section{Introduction}\label{sec:intro}
Gamma-ray bursts (GRBs) are the universe's most energetic 
electromagnetic events, with observed energies of 
between $10^{49}$ and $10^{54}$ ergs per burst occurring 
on timescales ranging from tens of milliseconds to hundreds of 
seconds \citep{zha11,gr13,kz15,am16}. Upon
correcting for relativistic beaming and large Lorentz factors,
these energies are narrowly distributed 
around $10^{51}$ ergs per burst \citep{napi17}.

In the 1980s a phenomenological GRB classification scheme 
was introduced to subdivide GRBs into Long and Short
classes \citep{maz81,nor84} primarily on the basis of duration. 
Subsequent observations provided by BATSE (the Burst And Transient 
Source Experiment) supported this division (e.g. \cite{kou93,kos96}); 
the two classes were modeled by 
overlapping lognormal duration distributions with a 
delineation occurring at roughly 2$s$. 
This simple taxonomy unfortunately
became ensconced in the literature before 
more robust statistical and machine learning techniques could 
be applied, and the general validity of the two-class results
overshadowed later concerns about the robustness of the
classification approach.

Today it is  widely accepted that Short and Long GRB
classes have 
different progenitors, and that they represent dissimilar 
physical phenomena \citep{nor01,bal03,zha09,luli10,luliz10,li16}. 
The large luminosities and short emission timescales of all GRBs
have been explained by models involving 
black hole formation\citep{woosley17,feng18,fqkc18,SongLiu18}. 
The stellar core collapse model used to explain long GRBs occurred on a timescale too long to explain short GRBs, so that models involving compact object mergers in binary systems were developed to explain these \citep{paczy86,usov92,peram10,berger14}.

The host galaxies and redshift distributions 
of Short GRBs and Long GRBs differ \citep{berger14,levan16},
with the more luminous Long GRBs being found in star-forming galaxies.
Some low-luminosity Long GRBs have been associated with Type Ic 
supernovae (SN) \citep{hjor03,camp06,pian06,blan16}, supporting
the idea that the Long GRBs in general are related to deaths of 
massive stars \citep{woo93,paczy98,wb06,blan16}.
For Short GRBs the absence of SN association, the location of these events
in metal-poor regions, and their lower luminosities disfavor a massive star origin
and point to compact binary mergers 
\citep{paczy86,usov92,berger14}.

Despite this, the 2$s$ boundary separating Long GRBs and Short GRBs is 
not robust: the position of the ''boundary`` between Long and Short GRBs
depends not only on the methodology used and the size of the database,
but also on instrumental characteristics such as the detector threshold
and spectral response (e.g. \cite{zha12,qin13}). Furthermore, some 
GRBs are not amenable to the simple classification scheme.
For example, GRB060614 is a long duration burst that
shares enough properties with Short GRBs for \cite{zzl07} to suggest 
a compact star merger origin. Similarly, the observed 
properties of the GRB 090426 indicate that this 1.24$s$ 
duration burst had a collapsar origin \citep{lu14,li16}.

Perhaps because of the lack of a clear 
demarcation between Short GRBs and Long GRBs, 
\cite{zha06} proposed classifying GRBs
based on their expected behaviors, as predicted from models involving
compact objects (Type I) or massive stars (Type II).
Subsequently, \cite{zha09} suggested that multi-wavelength 
properties could be used in conjunction with theoretical models
to characterize real physical GRB classes. 
\citet{li16} carried out a multi-wavelength study between 
duration-defined long-short and the Type I-Type II GRBs. 
They found several observables to be useful, but not
fully reliable, for classification, due to overlap between 
the populations.

The simple two-class interpretation is further complicated by
the statistical existence of a third GRB class.
Using multi- and uni-variate statistical analysis techniques,  
\cite{muk98} and \cite{hor98} found evidence for a third GRB
class in data from the Third BATSE Catalog \citep{m6}. 
Many authors \citep{hak00,bala01,rm02,hor02,hak03,bor04,hor06,chat07,zito15} 
have since confirmed the existence of this Intermediate 
GRB class in the same database
using statistical techniques and/or data mining algorithms. 
The Intermediate class has also been found in the Beppo-SAX \citep{hor09} 
and Swift data \citep{hor08,huja09,hor10,ht16}, even though
Beppo-SAX had a smaller effective area than BATSE, and
Swift works in a different energy range. 

The growing number of bursts detected by the Fermi Gamma-ray Burst Monitor (GBM) 
provides additional data on which GRB classification
schemes can be tested. GBM has a spectral energy response that is
similar to, but broader, than BATSE, and a surface area that is
smaller than that of BATSE. Given the complementary, yet different, characteristics of the 
Fermi GBM instrument to BATSE, Swift, and Beppo-SAX, the time has come 
to apply statistical clustering techniques to explore GRB classification using this instrument. 

The paper is organized as follows. Section 2 discusses
the properties of the Fermi GBM catalog, Section 3 defines
thirty six potential classification parameters and their structures, and 
Section 4 describes the classification process using sixteen 
GRB parameters following the elimination of sixteen duplicative
parameters. Section 5 discusses the results and Section 6 
provides the paper's conclusions.

\section{Classification Variables from the Fermi GBM Catalog}
On April 18, 2017, the Fermi GBM Catalog contained 2060 GRBs.\footnote{https://heasarc.gsfc.nasa.gov/W3Browse/fermi/fermigbrst.html}
The most recent GRB at that time was GRB170416583, although
the last GRB for which spectral fits were available was GRB170131969 (the 2011th GRB).
More than three hundred variables that might be used in GRB
classification are published for each GRB, including 
negative and positive uncertainties. Upon excluding the uncertainties, 
the number of available variables is less than two hundred. 
Most of these are spectral fit parameters, including 
fitting parameters for each of four different spectral models: 
Band, Comptonized (Comp), Power Law, and Smoothly 
Broken Power Law (SBPL); for details see \cite{3gbm16}. 
All four fits are applied to both the set of photons observed 
in the one second-peak (the 1024 ms peak flux) 
and the set of all photons in the burst (the fluence). 
 
Published analyses have used a variety of different
variables in their classification approaches. 
Some analyses use only the duration information  
\citep{hor98,bala01,rm02,hor02,hor08,huja09,hor09,zito15,
tarno15AA,tarno15ApSS,ht16,tarno16MNRAS,kulk17}, 
others primarily use the duration-hardness plane 
\citep{hor04,hor06,veres10,hor10,kb12,qin13,tsu14,sn15,rm16,yzj16,zycc16}, 
while still others use more than two variables 
\citep{muk98,hak03,chat07,kk11,lu14,li16,2017arXiv170305532Modak,chat17}.

For our analysis we have chosen to use thirty six
potentially useful classification variables, even if they
overlap in content, with the assumption that any deemed 
to be nonsense parameters can be removed later. 
There are two duration measures (T90 and T50),
two fluence measures (one based on Fermi GBM energy
channels and the other based on BATSE energy channels),
six peak flux measures (on the 64-ms, 256-ms, and 
1024-ms timescales measured in both the Fermi GBM and
BATSE energy channels), eight Band spectral fit
parameters ($\alpha$, $\beta$, $E_{\rm peak}$,
and the fit amplitude obtained from both the peak flux
and fluence spectra), eight broken power law
spectral fit parameters (low-energy index, high-energy
index, break energy, and fit amplitude obtained from
both the peak flux and fluence spectra), four single
power law spectral fit parameters (power law
index and the fit amplitude obtained from both the
peak flux and fluence spectra), and six Compton
spectral fit parameters (spectral index, peak
energy, and fit amplitude obtained from both the
peak flux and fluence spectra).
Because most of these parameters span large dynamical 
ranges, we use the base 10 logarithmic measures 
of these variables (lg) instead of the measures themselves. 
%
There is no need to take the logarithmic values  of the spectral indices
since these have already been obtained as power law functions.

\section{Statistical Clustering of GBM Data}\label{sec:GRB spatial distribution}
   
\subsection{Errors and correlations of pre-selected classification variables}

\begin{table}[t]\begin{center}
    \centering
    \caption{Numbers of bursts where the peak energy uncertainties are larger than 1500 keV, 800 
    keV, and 400 keV.}
	\begin{tabular}{|p{3cm}||p{1.3cm}|p{1.1cm}|p{1.1cm}|}\hline 
\backslashbox[3.4cm]{Epeak} {error} & \multicolumn{3}{c|}{\begin{tabular}{c|c|c} 
  $>1500$   &   ~$>$ 800~  &  $>$ 400   \\
\multicolumn{3}{c}{keV}\\  
\end{tabular}}   \\ 
\hline \hline 
FlncCompEpeak & 	65 & 	95 & 	145\\ \hline
FlncBandEpeak	 & 26 & 	49 & 	94\\ \hline
PflxCompEpeak & 	112 & 	155 & 	244\\ \hline
PflxBandEpeak	 & 54 & 	90	 & 160\\ \hline
PflxSbplBrken & 	253 & 	329 & 	429\\ \hline
FlncSbplBrken	 & 192	 & 246	 & 308\\ \hline
	\end{tabular}
	\label{tab:err2}
\end{center}
\end{table}
Large numbers of objects and many classification variables do not 
guarantee that a GRB classification will be successful.
Because GRBs are observed in low signal-to-noise regimes where
measurements are difficult to make with accuracy, large uncertainties
often accompany potential classification variables. 
Furthermore, the information contained in each classification variable 
is not always independent; many variables have content that
overlaps with that found in others ({\it e.g.,} T90 and T50). 

Before applying classification tools to this GRB sample,
we reduce the scatter in the data by removing 
GRBs having large measurement uncertainties; the numbers of such GRBs
are found in Tables ~\ref{tab:err2} - ~\ref{tab:err3}.
Among the 2060 GRBs in our sample, 1729 have usable 
spectral fits of one form or another and
1487 have values measured for all 36 variables.

\begin{table}[t]\begin{center}
    \caption{Numbers of bursts for which flux / fluence uncertainties exceed 100\% and 50\% of the observed values.}
	\begin{tabular}{|l||p{2cm}|p{2cm}|}\hline 
		 & \multicolumn{2}{c|}{Number of bursts with}     \\ 
		 & \multicolumn{2}{c|}{errors exceeding}    \\ 
		 &   \multicolumn{1}{c|}{100\%}  &   \multicolumn{1}{c|}{50\%} \\
		 \hline \hline 
T90		  & 147	  & 309 \\ \hline
T50		  & 127  & 	329\\ \hline
Fluence		  & 0	  & 1\\ \hline
FluenceBATSE	  & 0  & 	2\\ \hline
Pflux1024	  & 1	  & 5\\ \hline
Pflux64		  & 3  & 	13\\ \hline
Pflux256	  & 1	  & 1\\ \hline
Pflux1024BATSE	  & 0  & 	7\\ \hline
Pflux64BATSE	  & 1	  & 54\\ \hline
Pflux256BATSE	  & 0  & 	9\\ \hline
PflxPlawAmp	  & 0	  & 0\\ \hline
PflxCompAmp	  & 99  & 	456\\ \hline
PflxBandAmp	  & 377	  & 711\\ \hline
PflxSbplAmp	  & 6  & 	11\\ \hline
FlncPlawAmp	  & 0	  & 0\\ \hline
FlncCompAmp	  & 10  & 	92\\ \hline
FlncBandAmp	  & 238	  & 440\\ \hline
FlncSbplAmp	  & 1	  & 1\\ \hline
	\end{tabular}
	\label{tab:err1}
\end{center}
\end{table}

\begin{table}[t]\begin{center}
    \hfill{}
    \caption{Numbers of bursts where the spectral index uncertainties exceed 3, 1, and 0.5.}
	\begin{tabular}{|l||r|r|r|}\hline 
		        & error $>$ 3   & error $>$ 1   & err $>$ 0.5  \\ \hline \hline
flncbandalpha   & 45            & 	133         & 278\\ \hline
flncbandbeta	& 446           & 	589         & 740\\ \hline
pflxplawindex   & 0             & 	0           & 0\\ \hline
pflxcompindex	& 1             & 	45          & 248\\ \hline
pflxbandalpha   & 79            & 	241         & 516\\ \hline
pflxbandbeta    & 643           & 	836         & 1050\\ \hline
pflxsbplindx1   & 91            & 	208         & 361\\ \hline
pflxsbplindx2	& 439           & 	690         & 942\\ \hline
flncplawindex	& 0             & 	0           & 0\\ \hline
flnccompindex	& 2             & 	12          & 48\\ \hline
flncsbplindx1   & 56            & 	112         & 206\\ \hline
flncsbplindx2	& 239           & 	436         & 657\\ \hline
	\end{tabular}
	\hfill{}
	\label{tab:err3}
\end{center}
\end{table}

\begin{figure}[h!]\begin{center}
 \resizebox{\hsize}{!}{\includegraphics[width=15.61cm,
 height=9.5cm,angle=0]{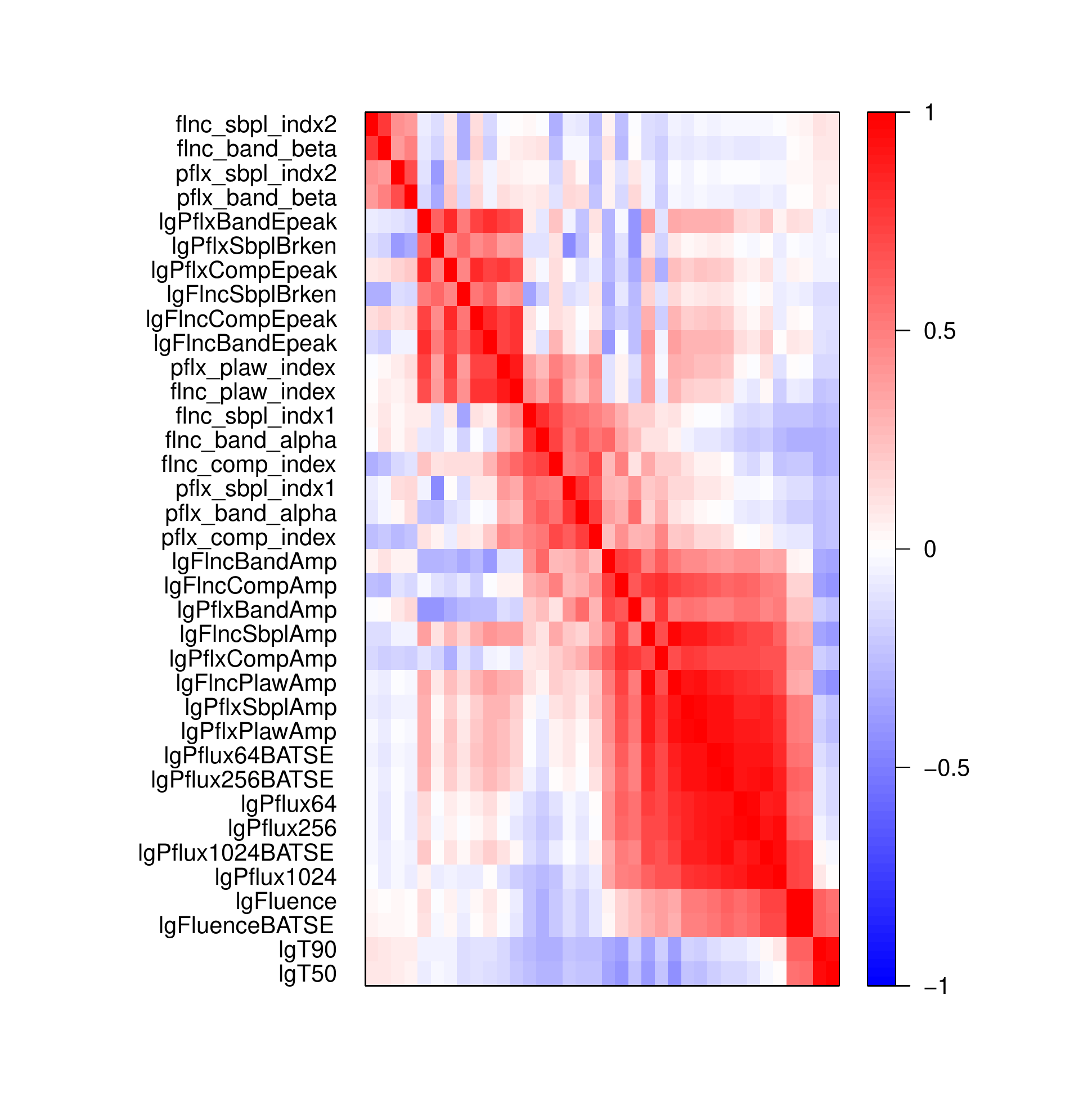} }
 
 \caption{\small{Spearman correlation matrix of 36 potential classification variables, 
                             seriated using the PCA\_angle method.}}
  \label{fig:36v1487}
\end{center}
\end{figure}

We have calculated the correlation matrix in order to determine the amount 
of information overlap found in the 36 classification variables.
The correlation matrix is found with the built-in R-environment \citep{rlang} \texttt{cor()} 
function using the available Pearson, Kendall, and Spearman methods. 
Seriation is performed on each of these matrices
with the package \texttt{seriation} \citep{seriation} using the function \texttt{seriate()} and the methods 
PCA and PCA\_angle. The results obtained from the different
techniques are very similar to one another, and the correlation matrix
constructed with the Spearman method and seriated with the 
PCA\_angle method is shown in Figure ~\ref{fig:36v1487}.

Clusters of variables are observed along the main diagonal
where content overlap is significant: 
The two fluences are related to T90 and T50 and to
the peak fluxes and the spectral amplitudes (the large 14x14
square in the lower right side of the matrix). The content overlap 
between fluence, peak flux, and duration make intuitive sense
as fluence is the time-integrated flux, measured over the burst's duration
({\it e.g.,}\cite{hak03}). Spectral amplitudes are also types of peak fluxes, 
but that are binned spectrally rather than temporally.
Low energy spectral indices form another cluster in the matrix, as 
these indices all measure low-energy burst behaviors.
The power law spectral indices overlap with
break energies rather than with the other spectral indices
because power-law indices identify single-component
spectral models that increase through measured energies 
rather than decreasing at high energies (as more complex spectral models do).
Finally, on the upper left side of the matrix, the four high energy
spectral indices form a distinct group.

\subsection{Spectral fits for GRBs in the Fermi GBM catalog}

\begin{table}[t]\begin{center}
    \hfill{}
    \caption{The number of GRBs having the best spectral model fit for fluence data (1802 GRBs) and for peak flux data (1792 GRBs).}
	\begin{tabular}{|p{20mm}||r|r|r|r|r|}
	\hline 
\backslashbox[24mm]{p flux}{flu} & PLaw & Comp  & Band & SBPL  & Total  \\ \hline \hline 
PLaw & 408 &  492  & 12 & 16 & 928 \\ \hline
Comp & 22 &  595  & 95  & 64 & 776 \\ \hline
Band & 0 &  4  & 47 & 9  & 60 \\ \hline
SBPL & 0 &  5  & 12 &  11 &  28 \\ \hline
no fit & 10 &    &  &   & 10 \\ \hline
Total  & 440 &  1096  & 166 & 100  &  1802 \\ \hline
	\end{tabular}
	\hfill{}
	\label{tab:spfit}
\end{center}
\end{table}

The Fermi GBM catalog contains information
about which of the four spectral fits is best for each burst. 
Spectral fits have been obtained at the time of the peak flux for
1792 of the 2060 GRBs, and using the fluence for
1802 of the 2060 GRBs.
Among the peak flux spectral fits, the power-law function provides the best fit
for 928 GRBs, the Compton model for 776 GRBs,
the Band function for 60 GRBs, and the SBPL function
for the remaining 28 GRBs. 
Among the fluence spectral fits, the power law model provides the best fit
for 440 GRBs, the Compton model for 1096 GRBs,
the Band model for 166 GRBs, and the SBPL model
for the remaining 100 GRBs.

Table ~\ref{tab:spfit} summarizes the consistency of
different spectral fit models obtained from fluence and peak flux spectral data,
with fits to peak flux data listed in rows and fits to fluence data listed in columns.
The peak fluxes of 10 bursts provide insufficient photon counts to
generate any spectral fit and are shown in the first column; the fluences of these 
bursts have only marginally better statistics resulting in simple power law fits.

\subsection{Revisiting the effectiveness of the spectral fit variables}

The spectral variables described in Section 3.1 have been included in our tabulations
regardless of the quality of these fits. 
To more accurately evaluate the potential effectiveness of the 
variables, we examine the reduced 
$\chi ^2$s of each fit, as published in the Fermi GBM catalog. 

\begin{figure}[h!]\begin{center}
 
 \resizebox{\hsize}{!}{\includegraphics[angle=0]{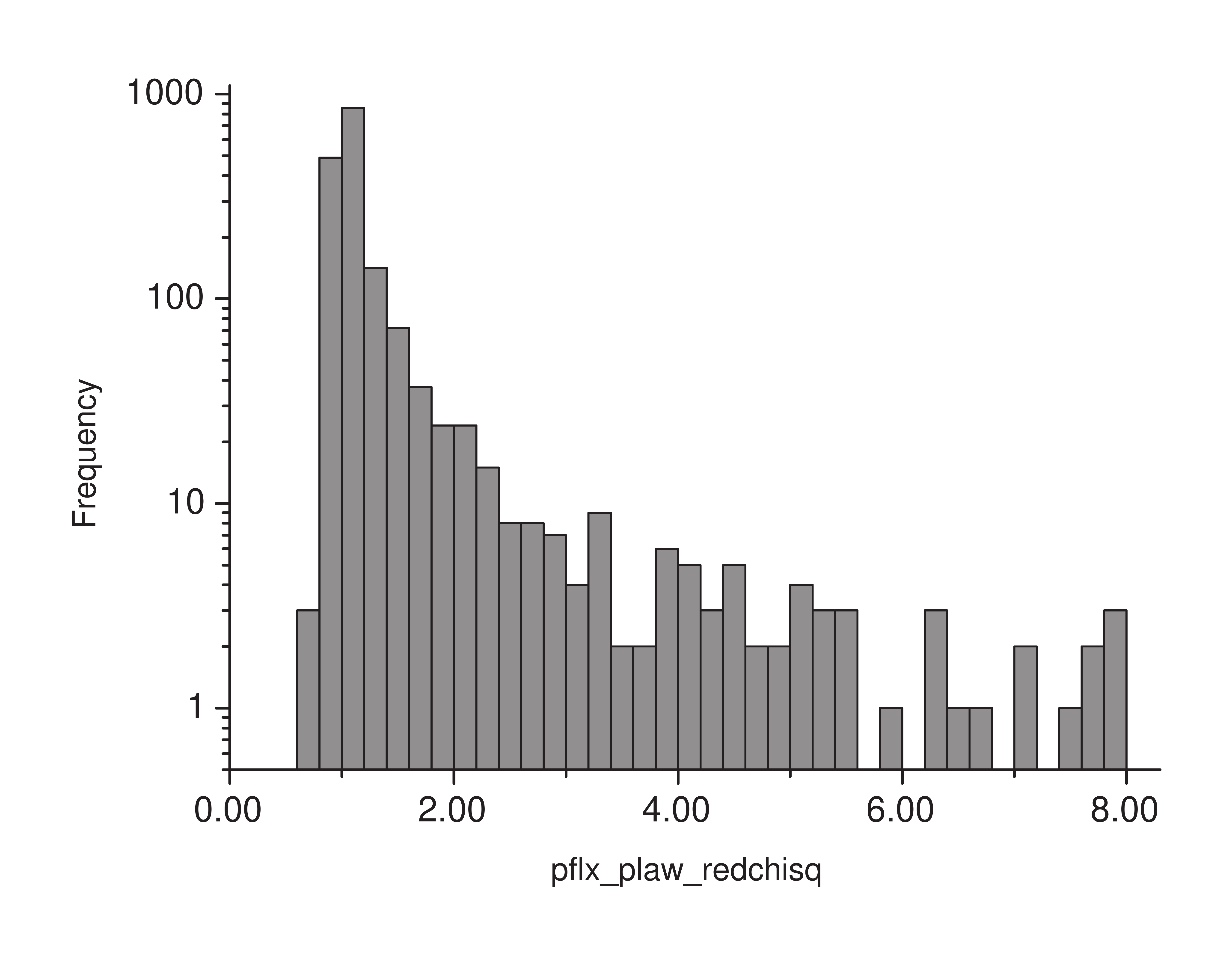} }
 
 \caption{\small{Reduced $\chi ^2$ distribution of power law spectral fits, obtained from peak flux spectra.}}
  \label{fig:pflxplawredchisq}
\end{center}
\end{figure}

\begin{figure}[h!]\begin{center}
 \resizebox{\hsize}{!}{\includegraphics[angle=0]{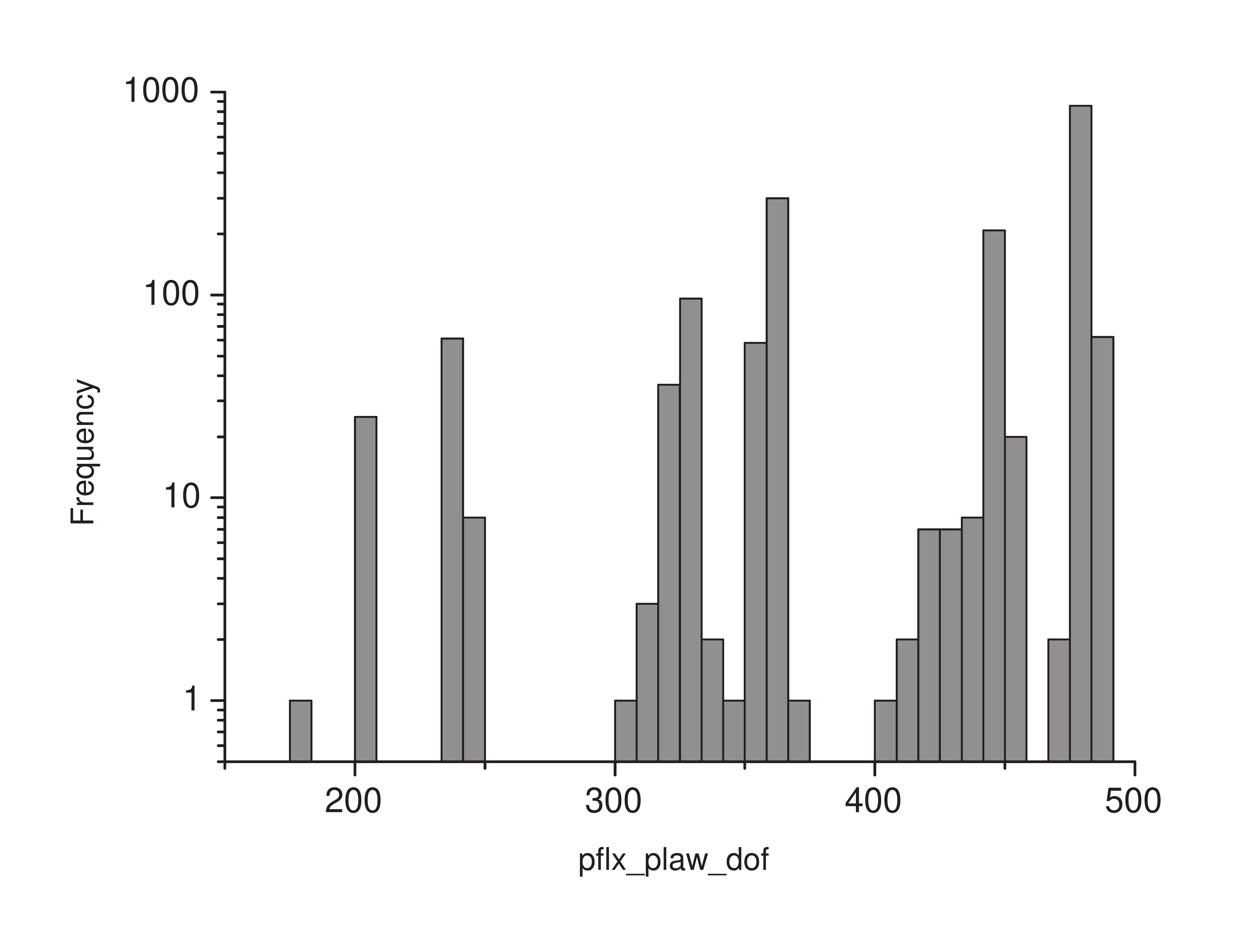} }
 \caption{\small{DOF distribution of power law spectral fits, obtained from peak flux spectra.}}
  \label{fig:dof}
\end{center}
\end{figure}

Figures ~\ref{fig:pflxplawredchisq} and ~\ref{fig:dof} demonstrate the 
reduced $\chi ^2$ and associated degree of freedom (DOF) 
distributions for the power law spectral fits obtained from peak flux measurements. 
We note that the DOF distribution is identical for all eight spectral fits.

\begin{figure}[h!]\begin{center}
 \resizebox{\hsize}{!}{\includegraphics[angle=0]{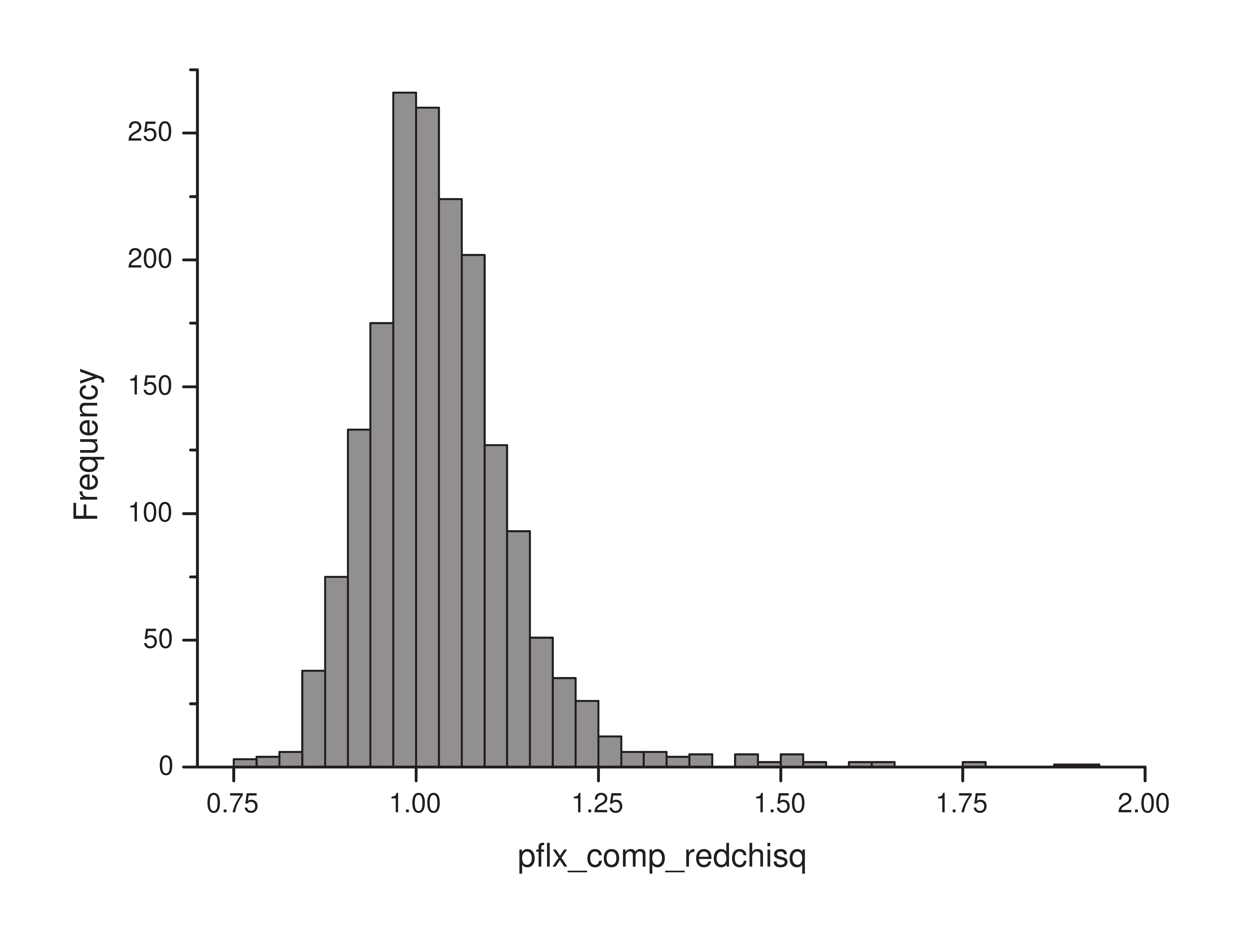} }
 \caption{\small{Reduced $\chi ^2$ distribution of Compton spectral fits, obtained from peak flux spectra.}}
  \label{fig:pflxcompredchisq3}
\end{center}
\end{figure}

Most of the DOF values exceed 430, 96\% are larger than 300, and 
only one is less than 200. For DOF = 400, a reduced $\chi ^2$ value 
of 1.172 indicates a 99\% significance (a value of 1.232 corresponds 
to 99.9\%). For DOF = 300, the reduced $\chi ^2$ value of 1.199 
indicates a 99\% significance (1.271 corresponds to 99.9\%), and for 
DOF = 200 the reduced $\chi ^2$ value of 1.247 indicates a 99\% 
significance (1.337 corresponds to 99.9\%). This means that, for the 
DOF values found in this distribution of Fermi GBM bursts, reduced 
$\chi ^2$ values of 1.25 indicate poor spectral fits and values of 1.35 
indicate very poor spectral fits. 
The reduced $\chi ^2$ values of the different spectral fits range from
being mostly poor ($> 1.25$) in the case of the power law model fits (see 
Figure ~\ref{fig:pflxplawredchisq}) to being generally good ($< 1.25$) in the 
case of the Compton model fits (see Figure ~\ref{fig:pflxcompredchisq3}).

\begin{figure}[h!]\begin{center}
 \resizebox{\hsize}{!}{\includegraphics[angle=0]{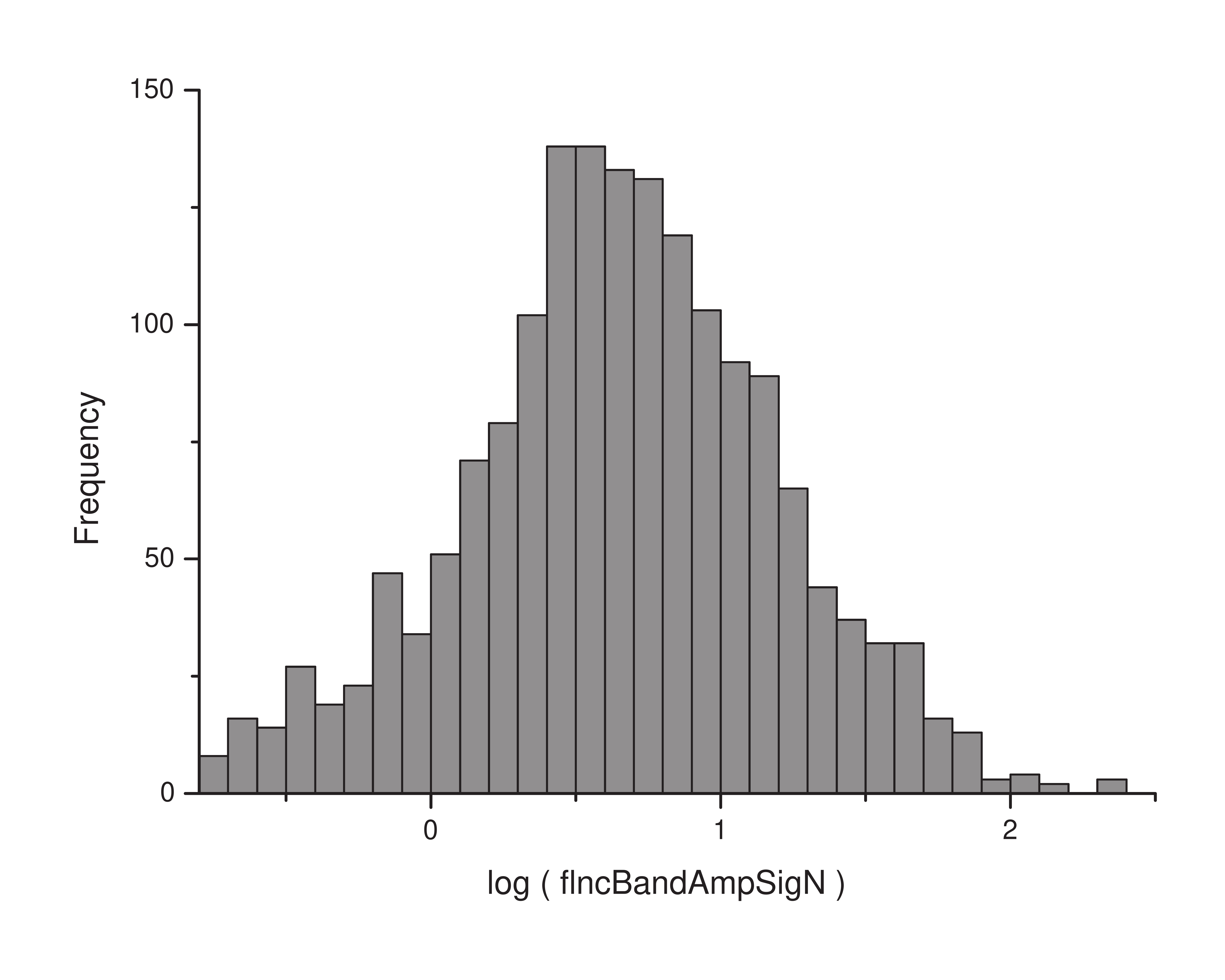} }
 \caption{\small{$S/N$ distribution of the Band spectral fit amplitude, obtained from fluence spectra.}}
  \label{fig:lncBandAmpSigN}
\end{center}
\end{figure}

The signal-to-noise ratio ($S/N$ = value/error) is another quality indicator 
that should be examined when considering the value of potential
classification variables.  The effects of $S/N$
on spectral fit measurements can be seen in
Figure ~\ref{fig:lncBandAmpSigN}, which shows the $S/N$ 
distribution of Band spectral fit amplitudes measured 
from fluence spectra. Several hundred GRBs in this distribution 
have ($S/N<1$), indicating that their fit amplitudes
are less than their fit amplitude uncertainties. Another few 
hundred GRBs have signal-to-noise ratios smaller than two or three.

\begin{figure}[h!]\begin{center}
 \resizebox{\hsize}{!}{\includegraphics[angle=0]{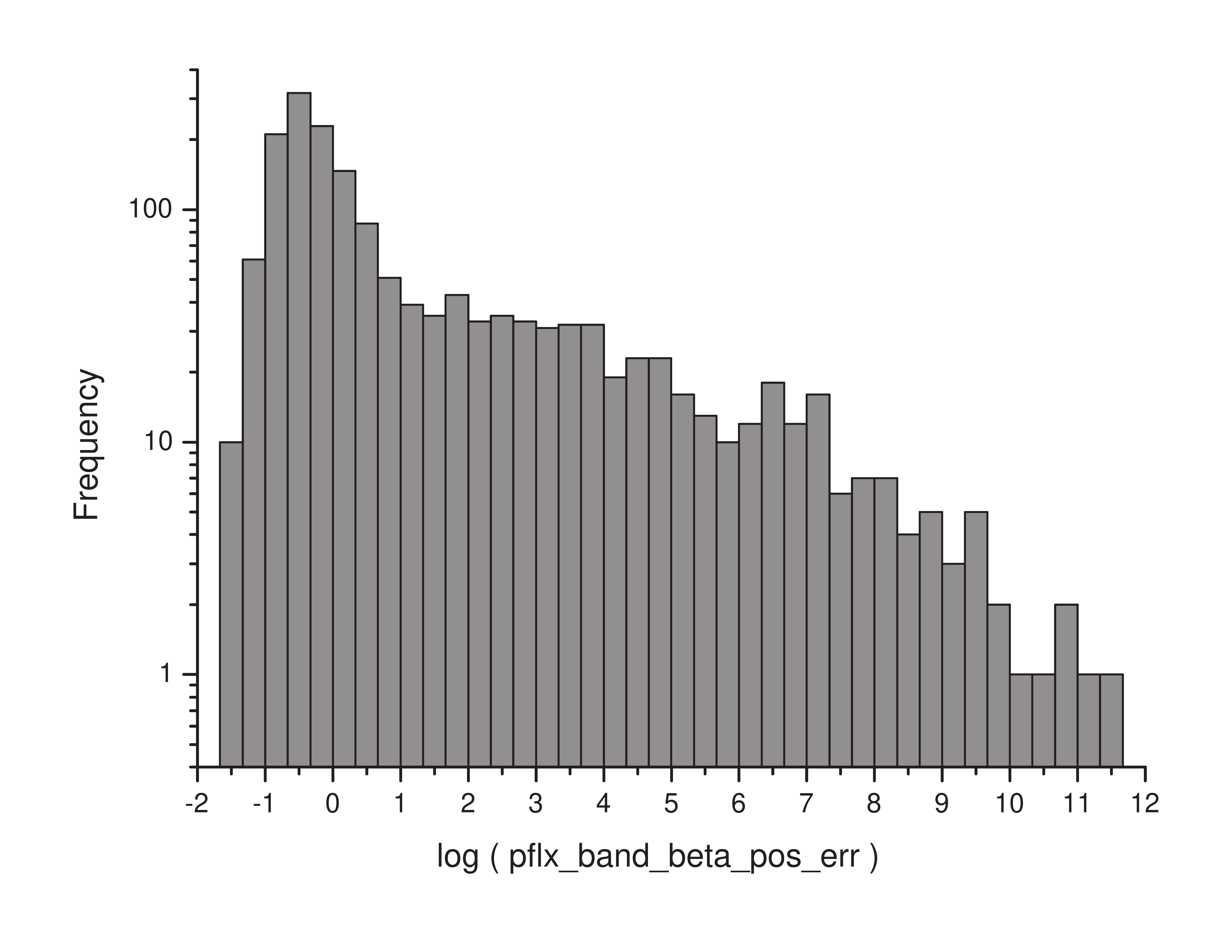} }
 \caption{\small{Band $\beta$ spectral indices, obtained from peak flux spectra.}}
  \label{fig:BandBetaPosErr}
\end{center}
\end{figure}

Very large uncertainties are also present in some of the high energy 
spectral indices measured for Fermi GBM bursts. For example, 
Figure ~\ref{fig:BandBetaPosErr} demonstrates the uncertainties
in the Band $\beta$ spectral index obtained from peak flux measurements. 
Since $\beta$ typically has values in a narrow range ($-3 \le \beta \le -2$), 
$\beta$ values having uncertainties ($\sigma_\beta > 1$)
are not likely to be useful for classification. Unfortunately, many
Fermi GBM bursts fitted by the Band model fall into this range; several
hundred have  $\sigma_\beta >10$, and many have $\beta$ values that 
exceed 100 and even 1000.

Due to the uncertainties in measuring high energy power-law spectral indices,
we have omitted the Band and SBPL fits (for all four cases, both fluence 
and peak flux fits) in this analysis.

We similarly analyze the $\chi ^2$ distributions of the Compton and power law
spectral models to determine whether the variables obtained from
these fits have potential value as classification variables.
Three hundred eighty one power law model 
and 60 Compton model $\chi ^2$ values of the 1792 peak flux spectra
are deemed unacceptable, as are 703 power law and 180 Compton model  
$\chi ^2$ values. 
If we want to use for classification the acceptable spectral model variables from
both peak flux and fluence spectral fits,
then only the Compton model produces enough variables for
a sufficiently large sample. This reduces the sample size by roughly half.  
We also exclude GRBs where the uncertainty in any variable exceeds 
50\% of that variable's value. This excludes, for example, 
184 GRBs on the basis of T90 and 288 GRBs on the basis of
peak flux amplitude. Only 803 GRBs satisfy all of our requirements
for a well-measured classification variable database. 
Finally, two more bursts have been excluded using the 
\texttt{HDoutliers()} function of the R-package \texttt{HDoutliers} \citep{hdout}.

The 801 remaining GRBs are chosen to be the sample for the analysis 
described in Section 4. The list of these GRBs and the values 
of the variables can be found at 
http://itl7.elte.hu/~hoi/grb/f1704C801v16.csv

\section{Cluster analysis with 16 variables}
\begin{figure}[ht!]\begin{center}
 \resizebox{\hsize}{!}{\includegraphics[height=3.1cm,width=4.91cm,
 angle=0]{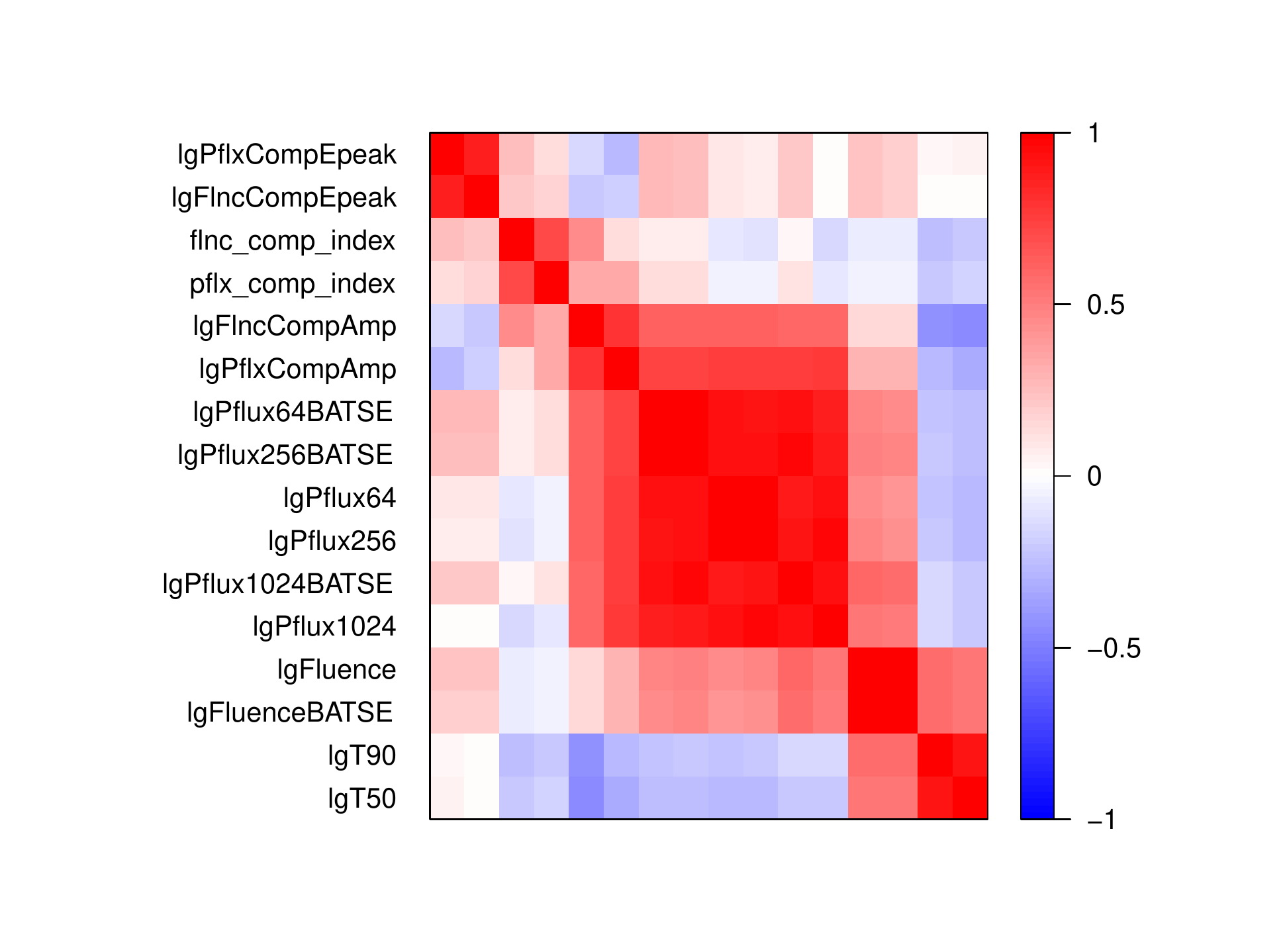} }
 
 \caption{\small{Spearman correlation matrix of 16 potential classification variables, 
                             seriated using the PCA\_angle method.}}
  \label{fig:PCAangle16}
\end{center}
\end{figure}

We have reduced the number of GRB classification variables by limiting our
sample to the following sixteen well-measured characteristics of bursts in the Fermi GBM Catalog
(note that all denoted by 'lg' represent logarithmic values): 
lgT90 (T90), lgT50 (T50), lgfluence (fluence), lgPflux64 (64 ms peak flux),
lgPflux256 (256 ms peak flux), lgPflux1024 (1024 ms peak flux), lgfluenceBATSE
(fluence in the BATSE energy channels), lgPflux64BATSE (64 ms peak flux in the
BATSE energy channels), lgPflux256BATSE, (256 ms peak flux in the BATSE energy
channels), lgPflux1024BATSE (1024 ms peak flux in the BATSE energy channels), 
lgflncCompAmp (Compton amplitude from fluence spectral fit), 
lgflncCompEpeak (Compton Epeak from fluence spectral fit),
flncCompIndex (Compton power law index from fluence spectral fit), 
lgpflxCompAmp (Compton amplitude from peak flux spectral fit), 
lgpPflxCompEpeak (Compton Epeak from peak flux spectral fit), and
pflxCompIndex (Compton power law index from fluence spectral fit). 
Using the techniques described in Section 3.1, 
we obtain the correlation matrix using the built-in \texttt{cor()} 
function using the Spearman method and seriate the results with the \texttt{seriate} function of 
the \texttt{FactoMineR} package \citep{FactoMine} via the PCA\_angle method. 
The result can be seen in Figure ~\ref{fig:PCAangle16}.

Four distinct blocks of correlated variables are apparent in the matrix: 

block1:  lgPflxCompEpeak and lgflncCompEpeak

block2: pflxCompIndex and flncCompIndex

block3: lgPflux64, lgPflux256, lgPflux1024, lgPflux\-64BATSE, lgPflux256BATSE, 
lgPflux1024BATSE, lgflncCompAmp, lgpflxCompAmp

block4: lgT90, lgT50, lgfluence, lgfluenceBATSE

\begin{table}[t]\begin{center}
    \caption{Eigenvalues of the PCA from 16 variables.}
	\begin{tabular}{|l||c|c|c|c|c|c|c|c|c|}\hline 
		   &  Eigenvalue   & cumulative \%  \\ \hline \hline 
PC1 &   7.468   & 46.7  \\ \hline
PC2 &    3.408  &  68  \\ \hline
PC3 &    2.458  &  83.3  \\ \hline
PC4 &    1.522   &  92.9 \\ \hline
PC5 &    0.411   & 95.4 \\ \hline
PC6 &    0.195  & 96.6 \\ \hline
	\end{tabular}
	\label{tab:6pc}
\end{center}
\end{table}

We have carried out PCA (principal component analysis) for the Fermi GBM bursts
using these 16 variables. 
Table ~\ref{tab:6pc} identifies the six largest PCs, while Table ~\ref{tab:evect}
shows the coefficients of the four largest PCA eigenvectors,
and Table ~\ref{tab:commu} identifies the communalities of these PCs.
To determine whether or not PCs are important, we use the
criterion that significant PCs should be larger than unity
and yield cumulative percentages larger than $80\%$ (see \cite{jol86book}).
Table ~\ref{tab:6pc} demonstrates that the top five PCs contain
a combined 95.2\% of the non-overlapping information contained in the
sixteen variables, and the top six PCs contain 96.6\% of
that information.

\begin{table}[t]\begin{center}
\scriptsize    
    \caption{The coefficients of the largest four PCA eigenvectors.}
	\begin{tabular}{|l||c|c|c|c|}\hline 
eigenvectors & 1st PC & 2nd PC  & 3rd PC & 4th PC  \\ \hline \hline 
pflxcompindex&0.109&-0.407&0.558&0.591\\ \hline
flnccompindex&0.060&-0.455&0.564&0.579\\ \hline
lgT90&-0.248&0.899&-0.033&0.257\\ \hline
lgT50&-0.288&0.870&0.012&0.292\\ \hline
lgFluence&0.549&0.753&0.163&0.228\\ \hline
lgFluenceBATSE&0.521&0.768&0.126&0.274\\ \hline
lgPflux1024&0.941&0.141&-0.208&-0.103\\ \hline
lgPflux64&0.949&-0.013&-0.089&-0.197\\ \hline
lgPflux256&0.961&0.028&-0.136&-0.181\\ \hline
lgPflux1024BATSE&0.963&0.112&0.031&-0.005\\ \hline
lgPflux64BATSE&0.961&-0.035&0.104&-0.109\\ \hline
lgPflux256BATSE&0.976&0.001&0.070&-0.082\\ \hline
lgPflxCompAmp&0.812&-0.234&-0.299&0.329\\ \hline
lgFlncCompAmp&0.695&-0.459&-0.166&0.384\\ \hline
lgFlncCompEpeak&0.189&0.025&0.886&-0.319\\ \hline
lgPflxCompEpeak&0.187&0.069&0.890&-0.310\\ \hline
	\end{tabular}
	\label{tab:evect}
\end{center}
\end{table}

\begin{table}[t]\begin{center}
    \caption{The communalities obtained using three and four PCs.}
	\begin{tabular}{|l||c|c|c|c|c|c|c|c|c|}\hline 
\backslashbox {variables} {communalities}	 & 3 PC & 4 PC   \\ \hline \hline 
pflxcompindex&	0.49&	0.839\\ \hline
flnccompindex&	0.53&	0.865\\ \hline
lgT90&	0.872&	0.939\\ \hline
lgT50&	0.842&	0.927\\ \hline
lgFluence&	0.897&	0.949\\ \hline
lgFluenceBATSE&	0.879&	0.955\\ \hline
lgPflux1024&	0.951&	0.961\\ \hline
lgPflux64&	0.91&	0.949\\ \hline
lgPflux256&	0.944&	0.977\\ \hline
lgPflux1024BATSE&	0.941&	0.941\\ \hline
lgPflux64BATSE&	0.937	&0.949\\ \hline
lgPflux256BATSE&	0.959&	0.966\\ \hline
lgPflxCompAmp	&0.805	&0.913\\ \hline
lgFlncCompAmp&	0.723	&0.871\\ \hline
lgFlncCompEpeak&	0.822&	0.925\\ \hline
lgPflxCompEpeak	&0.833&	0.93\\ \hline

	\end{tabular}
	\label{tab:commu}
\end{center}
\end{table}

\begin{figure}[ht]\begin{center}
 \resizebox{\hsize}{!}{\includegraphics[height=3.1cm,angle=0]{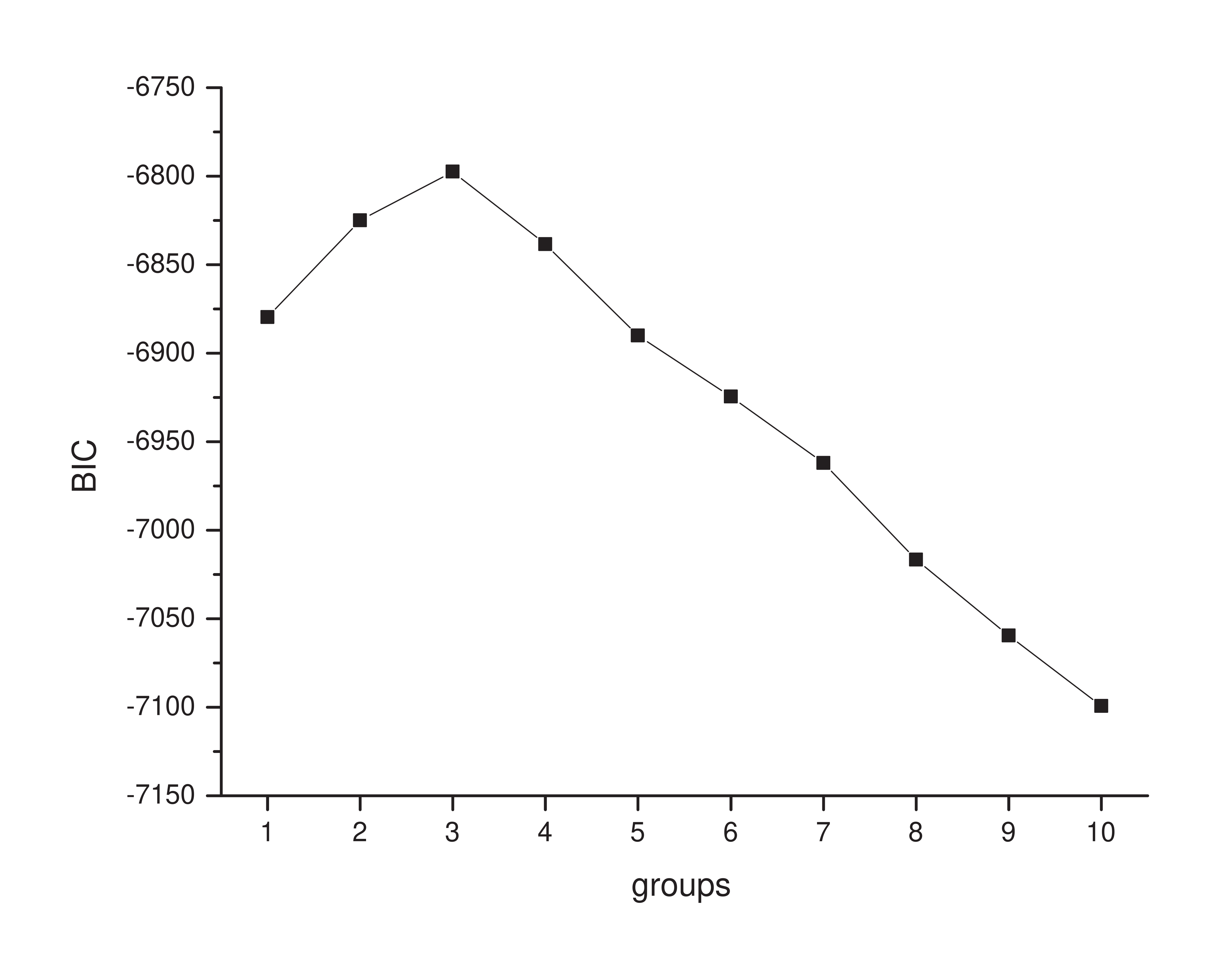} }
 
 \caption{\small{The Bayesian Information Criterion (BIC) suggests
that the three largest PCs extracted from the Fermi GBM 16-variable data
optimally describe three GRB classes}}
  \label{fig:bic}
\end{center}
\end{figure}

Table ~\ref{tab:commu} shows that the 
communalities obtained from three PCs are almost as robust as those obtained from four PCs
(although the communalities of the two Compton indices
are reduced to only 50\% with three PCs), so we choose
the three largest PCs for our analysis. 
The optimal number of classes contained within
these PCs can be found with guidance from the 
Bayesian Information Criterion (BIC) \citep{schw78}.
Figure ~\ref{fig:bic} shows that the maximum
BIC value (obtained from the \texttt{mclust()} function of the 
\texttt{mclust} \citep{NbClust} R-package) suggests
that the three PCs from the Fermi GBM 16-variable data
optimally describe three GRB classes.

Cluster analysis is performed on the 801 GRB sample
using the \texttt{mclust()} function with the 
aforementioned three PCs serving as classification variables.
The function returns three classes: Class 1 (containing 427 GRBs), 
Class 2 (containing 340 GRBs) and Class 3 (containing 34 GRBs). 
The values returned by the \texttt{mclust()} function also contain 
probabilities $p1_i$,$ p2_i$, and $p3_i$ that 
the $i$th GRBs 
belongs to Class 1, Class 2 or Class 3, respectively. 
We sum these probabilities to get $p1=400.1$ for Class 1, 
$p2=365.5$ for Class 2 and $p3=35.4$ for Class 3. These numbers differ from 
427, 340 and 34 because each GRB has been placed into the group
for which its probability is largest, and the non-integer counts of the
group elements created in this way does not necessarily yield the
same total as the some of the integer count probabilities. 

\begin{figure}[ht]\begin{center}
 \resizebox{\hsize}{!}{\includegraphics[height=3.1cm,angle=0]{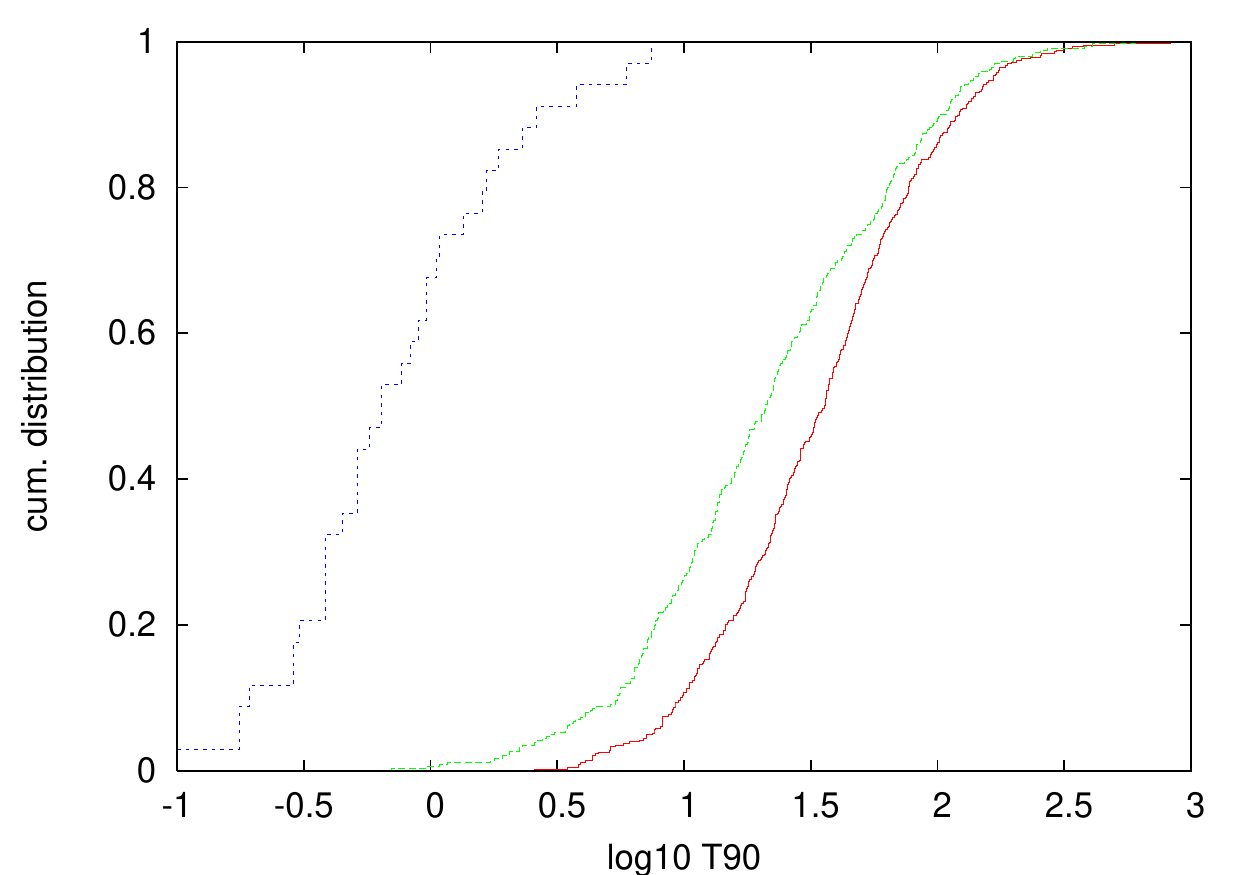} }
 
 \caption{\small{The $\log(T90)$ distributions of the three classes.
 Class 1 is shown in red, Class 2 in green, and Class 3 in blue.}}
  \label{fig:t90elo}
\end{center}
\end{figure}

\begin{figure}[ht]\begin{center}
 \resizebox{\hsize}{!}{\includegraphics[height=3.1cm,angle=0]{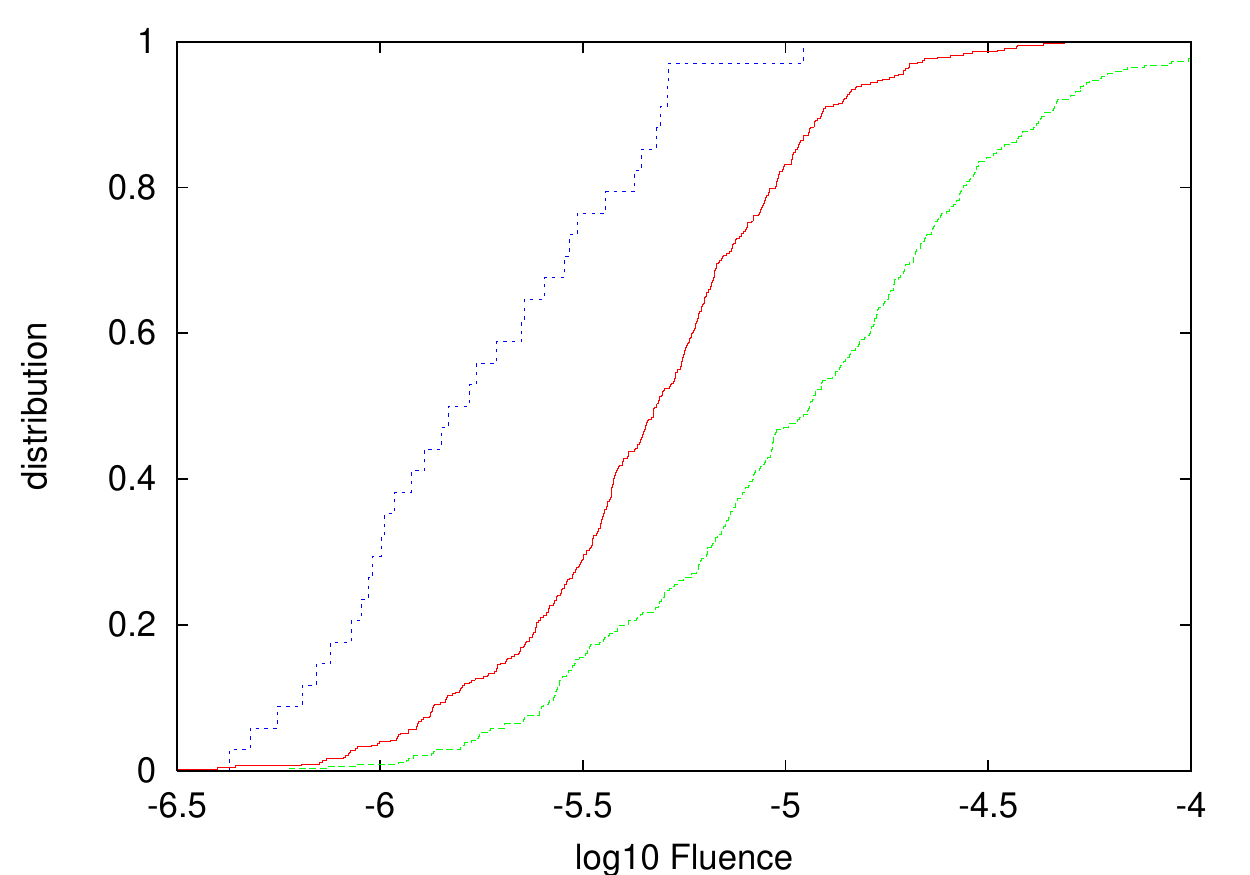} }
 
 \caption{\small{The fluence distributions of the three GRB classes.
 Class 1 is red, Class 2 is green and Class 3 is blue.}}
  \label{fig:flncelo}
\end{center}
\end{figure}

We have calculated the $\log(T90)$ distributions of all three GRB classes
by assuming that each GRB belongs to the class associated with
it largest cluster probability (Max[$p1_i$, $p2_i$, or $p3_i$]). 
The three $\log(T90)$ distributions 
are shown in Figure ~\ref{fig:t90elo}, after having been normalized 
by the factors 427 for Class 1, 340 for Class 2 and 34 for Class 3. 
In addition to duration, we have also calculated the 
class distributions for the other fifteen variables. 
The fluence distributions of the GRB classes 
can be seen in Figure ~\ref{fig:flncelo}. 
One can easily see from these figures that 
a short duration GRB class (Class 3) is present, consistent with
results obtained from other GRB experiments using a variety
of classification techniques.

\begin{figure}[ht!]\begin{center}
 \resizebox{\hsize}{!}{\includegraphics[height=3.1cm,angle=0]{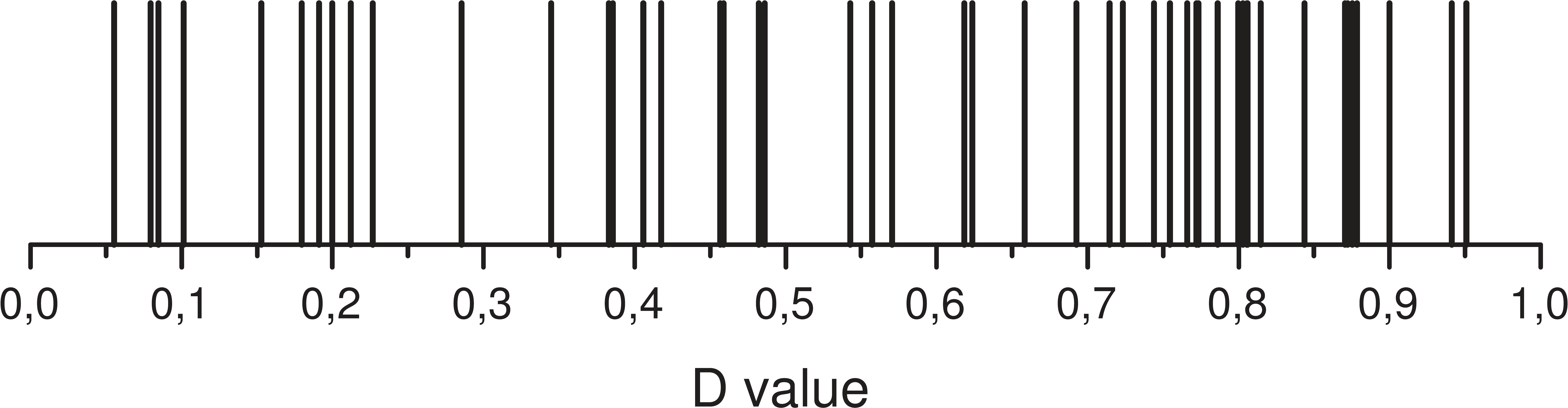} }
 
 \caption{\small{D values calculated from the group variable
   distributions.}}
  \label{fig:Dvalue}
\end{center}
\end{figure}

We have calculated the maximal differences between the three 
class distributions for each of the 16 observed parameters 
using the Kolmogorov-Smirnov statistic
(implemented in the R environment by the function \texttt{ks.test()});
the results are listed in Table ~\ref{tab:cumD} and are shown
in Figure ~\ref{fig:Dvalue}.
For the logT90 distributions, the maximal
distance between Class 3 and Class 1 is 0.95
and between Class 3 and Class 2 is 0.87,
indicating that the Class 3 durations are discernibly 
shorter than those of the other two classes. In contrast,
the maximal distance between Class 1 and Class 2 is only 0.212.

\begin{table}[t]\begin{center}
\caption{Maximal distances between classes (Class 1 = C1, Class 2 = C2, Class 3 = C3).}
	\begin{tabular}{|l||c|c|c|c|c|c|c|c|c|}\hline 
  &  \scriptsize{C1 v. C2}  & \scriptsize{C1 v. C3} & \scriptsize{C2 v. C3}  \\ \hline \hline 
pflxcompindex & 	0.101 & 	0.458 & 	0.417\\ \hline
flnccompindex	 & 0.055 & 	0.542 & 	0.57\\ \hline
lgT90	 & 0.212 & 	0.95 & 	0.87\\ \hline
lgT50	 & 0.226  & 	0.941 & 	0.9\\ \hline
lgFluence & 	0.382 & 	0.486 & 	0.723\\ \hline
lgFluenceBATSE	 & 0.385 & 	0.658	 & 0.805\\ \hline
lgPflux1024 & 	0.878 & 	0.344 & 	0.623\\ \hline
lgPflux64 & 	0.872	 & 0.772 & 	0.2\\ \hline
lgPflux256 & 	0.875 & 	0.743 & 	0.285\\ \hline
lgPflux1024BATSE & 	0.802 & 	0.456	 & 0.482\\ \hline
lgPflux64BATSE	 & 0.805 & 	0.843 & 	0.191\\ \hline
lgPflux256BATSE	 & 0.814 & 	0.754 & 	0.179\\ \hline
lgPflxCompAmp	 & 0.692 & 	0.618 & 	0.152\\ \hline
lgFlncCompAmp	 & 0.557 & 	0.786 & 	0.405\\ \hline
lgFlncCompEpeak	 & 0.085 & 	0.799 & 	0.773\\ \hline
lgPflxCompEpeak	 & 0.079 & 	0.765 & 	0.714\\ \hline
	\end{tabular}
	\label{tab:cumD}
\end{center}
\end{table}

\begin{figure}[ht!]\begin{center}
 \resizebox{\hsize}{!}{\includegraphics[height=3.1cm,angle=0]{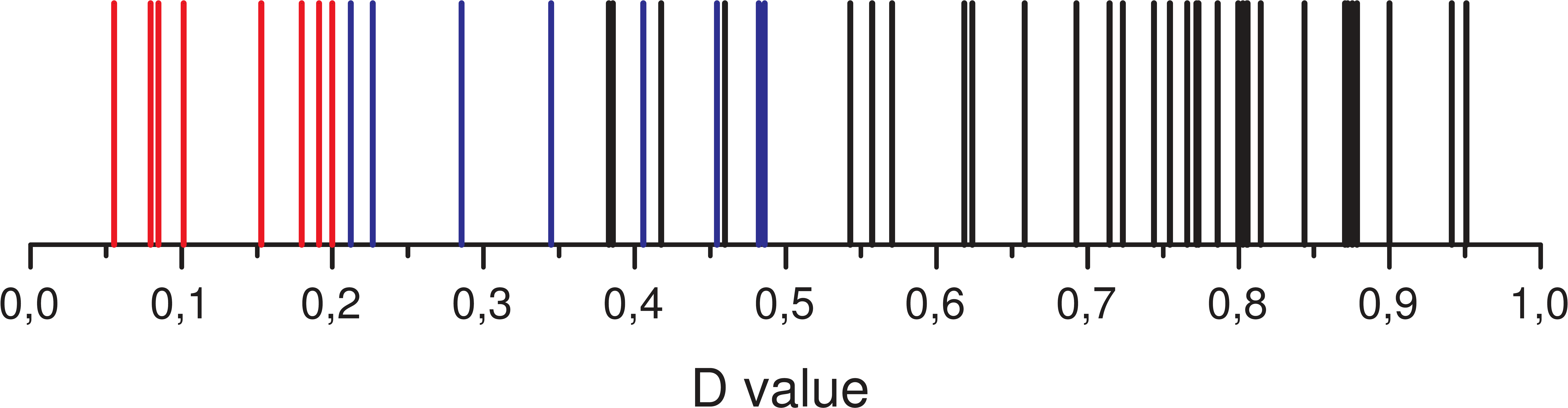} }
 
 \caption{\small{D values calculated from the class variable distributions. 
 Red is nonsignificant, blue may be significant by more than 99.9\%, and black is highly significant.}}
  \label{fig:Dszines}
\end{center}
\end{figure}

\begin{table}[t]\begin{center}
    \caption{The probability that the characteristics of two classes are similar. A probability greater than 0.01 indicates that two distributions are not significantly different.}
\small
	\begin{tabular}{|l||c|c|c|c|c|c|c|c|c|}\hline 
  &  \scriptsize{C1 v. C2}  & \scriptsize{C1 v. C3} & \scriptsize{C2 v. C3}  \\ \hline \hline 
pflxcompindex	&0.041&	0&	0\\ \hline
flnccompindex&	0.608&	0&	0\\ \hline
lgT90	& $10^{-7} $&	0&	0\\ \hline
lgT50	&$10^{-7} $&	0&	0\\ \hline
lgFluence&	0&	$10^{-6} $&	0\\ \hline
lgFluenceBATSE&	0&	0&	0\\ \hline
lgPflux1024&	0&	0.001&	0\\ \hline
lgPflux64&	0&	0	&0.168\\ \hline
lgPflux256&	0&	0&	0.013\\ \hline
lgPflux1024BATSE&	0&	$10^{-5} $&$10^{-6} $\\ \hline
lgPflux64BATSE&	0&	0&	0.209\\ \hline
lgPflux256BATSE&	0&	0&	0.273\\ \hline
lgPflxCompAmp	&0&	0&	0.464\\ \hline
lgFlncCompAmp	&0&	0&	$10^{-4} $\\ \hline
lgFlncCompEpeak	&0.133	&0	&0\\ \hline
lgPflxCompEpeak	&0.184&	0&	0\\ \hline

	\end{tabular}
	\label{tab:mcprob}
\end{center}
\end{table}

\begin{table}[t]\begin{center}
    \caption{Mean values of the sixteen GRB classification variables for each class. Within a row, black numbers 
    indicate a negligible difference, while blue numbers differ significantly. Red numbers differ in a highly significant manner 
    from the other two values.}
	\begin{tabular}{|l||c|c|c|c|c|c|c|c|c|}\hline 
  &   C1  & C2 & C3   \\ \hline \hline 
pflxcompindex& 	-0.7164& 	-0.7264& 	{\color{red}-0.344}\\ \hline
flnccompindex& 	-0.9761	& -1.0082& 	{\color{red}-0.566}\\ \hline
lgT90	& {\color{blue} 1.5343}& 	{\color{blue} 1.3345}& 	{\color{red} -0.1418}\\ \hline
lgT50& 	{\color{blue}1.1082}& 	{\color{blue}0.8387}	& {\color{red}-0.5146}\\ \hline
lgFluence& 	{\color{red}-5.3316}& {\color{red}-4.9571}& {\color{red}-5.7669}\\ \hline
lgFluenceBATSE& {\color{red}-5.5928}& {\color{red}-5.2326}& 	{\color{red}-6.184}\\ \hline
lgPflux1024& {\color{blue}0.6154}& {\color{red}1.1047}& {\color{blue}0.758}\\ \hline
lgPflux64& 	{\color{red}0.8054}& 	1.2518& 	1.2589\\ \hline
lgPflux256& 	{\color{red}0.7046}& 	1.1928& 	1.0924\\ \hline
lgPflux1024BATSE& {\color{blue}0.1862}& {\color{blue}0.7154}& {\color{blue}0.4185}\\ \hline
lgPflux64BATSE& 	{\color{red}0.4251}& 	0.8946& 	0.9353\\ \hline
lgPflux256BATSE	& {\color{red}0.3026}& 	0.8259& 	0.7733\\ \hline
lgPflxCompAmp& 	{\color{red}-1.7188}& 	-1.1881	& -1.2939\\ \hline
lgFlncCompAmp	& {\color{red}-2.1025}& {\color{blue}-1.7452}& {\color{blue}-1.5708}\\ \hline
lgFlncCompEpeak	& 2.3165& 	2.3008	& {\color{red}2.9907}\\ \hline
lgPflxCompEpeak	& 2.3663& 	2.351& 	{\color{red}2.9618}\\ \hline
	\end{tabular}
	\label{tab:mean}
\end{center}
\end{table}

\begin{figure}[ht!]\begin{center}
 \resizebox{\hsize}{!}{\includegraphics[height=3.1cm,angle=0]{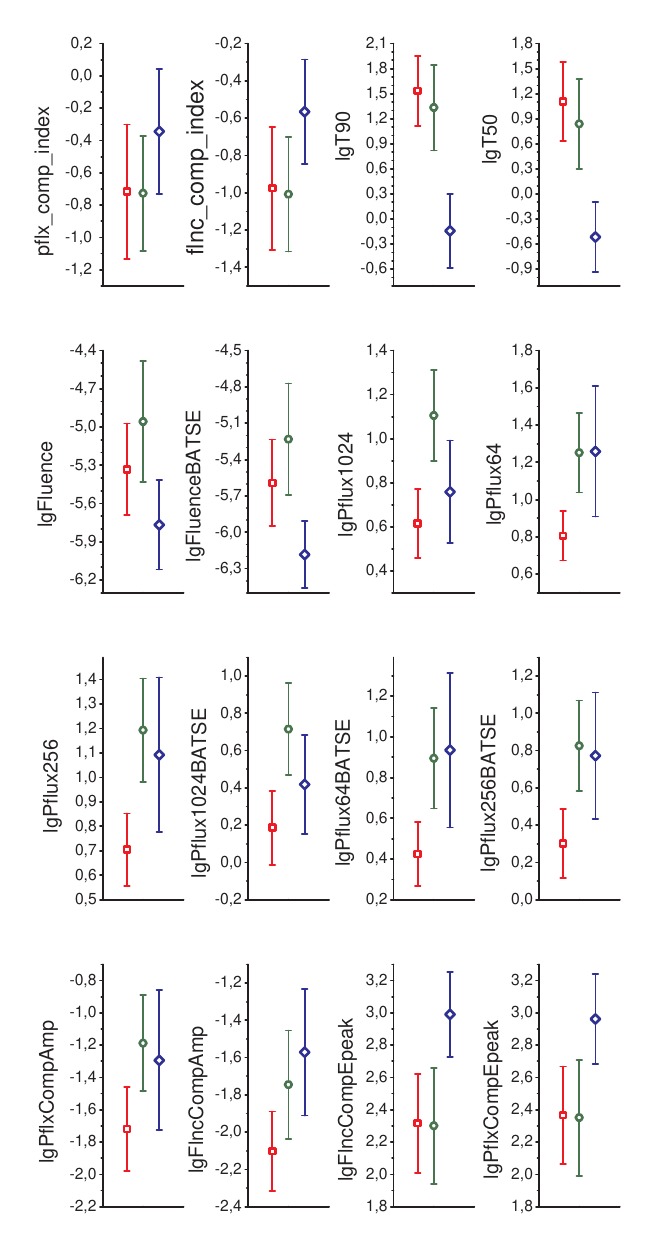} }
 
 \caption{\small{Mean values and their standard deviation of the 
 sixteen GRB classification variables for each class 
 (Class1 is red, Class2 is green and Class3 is blue).}}
  \label{fig:meansd}
\end{center}
\end{figure}

The KS probabilities associated with these maximal differences
are shown in Table ~\ref{tab:mcprob}. Using a significance 
requirement of at least 99.9\%, probabilities larger than 0.001 
indicate that there is no significant difference between 
two distributions. A near-zero probability therefore indicates a
very significant difference (In our calculation zero means less than $10^{-7}$).
Similar to Figure ~\ref{fig:Dvalue}, which plots the numbers
from Table ~\ref{tab:cumD}, Figure ~\ref{fig:Dszines} 
indicates the same D values colored according to
their probabilities. 
Red indicates no significant 
difference ($p > 0.1\%$), black
indicates a very significant difference, and blue means
that the two compared distributions may differ significantly.

\begin{figure}[ht!]\begin{center}
 \resizebox{\hsize}{!}{\includegraphics[height=4.6cm,width=4.91cm,
 angle=0]{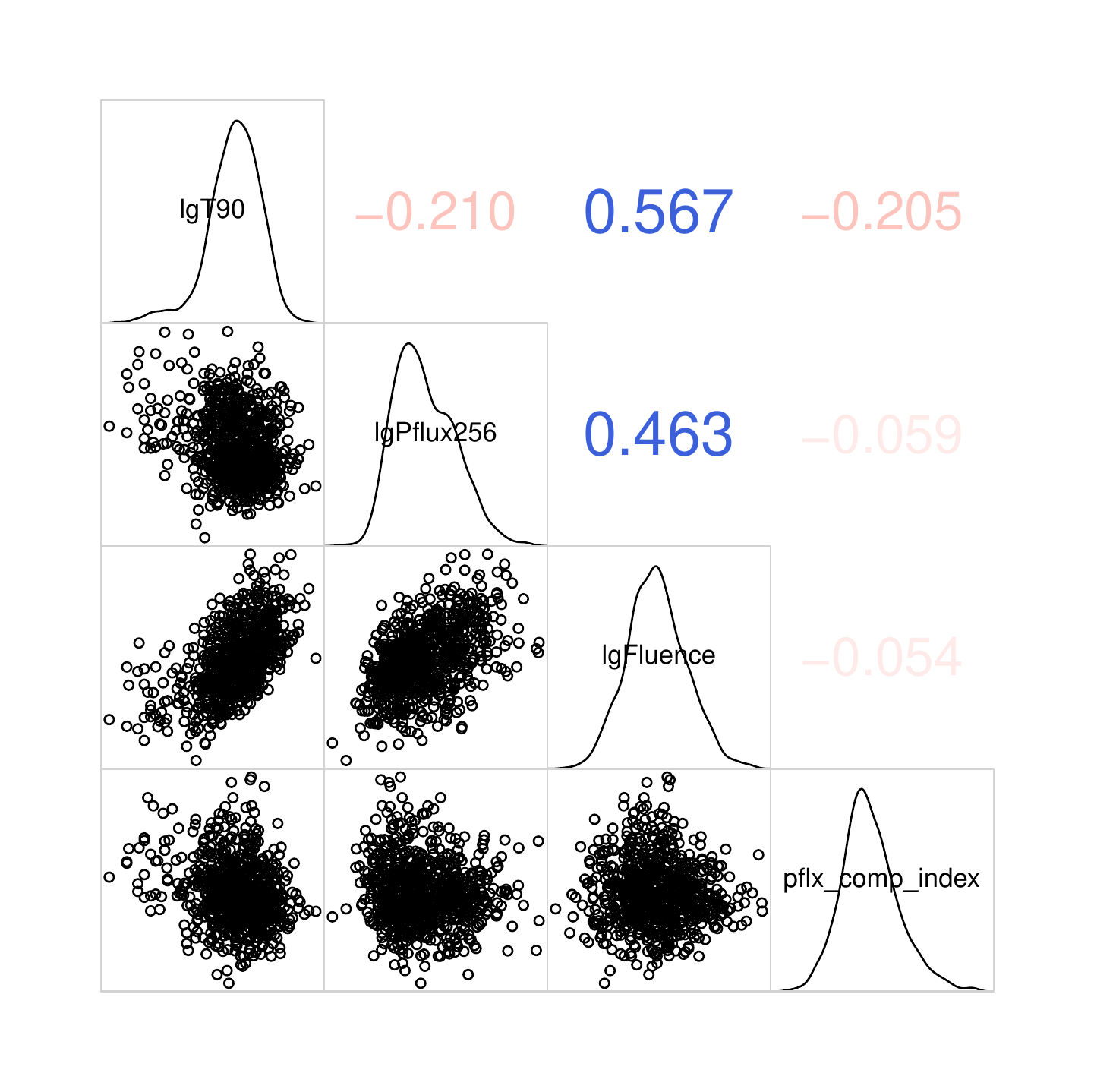} }
 \caption{\small{The 2-Dimensional distributions of lgT90, lgPflux256, 
 lgFluence and pflxcompindex 
 and their correlation coefficients.}}
  \label{fig:n2D1}
\end{center}
\end{figure}

\begin{figure}[ht!]\begin{center}
 \resizebox{\hsize}{!}{\includegraphics[height=4.6cm,width=4.91cm,
 angle=0]{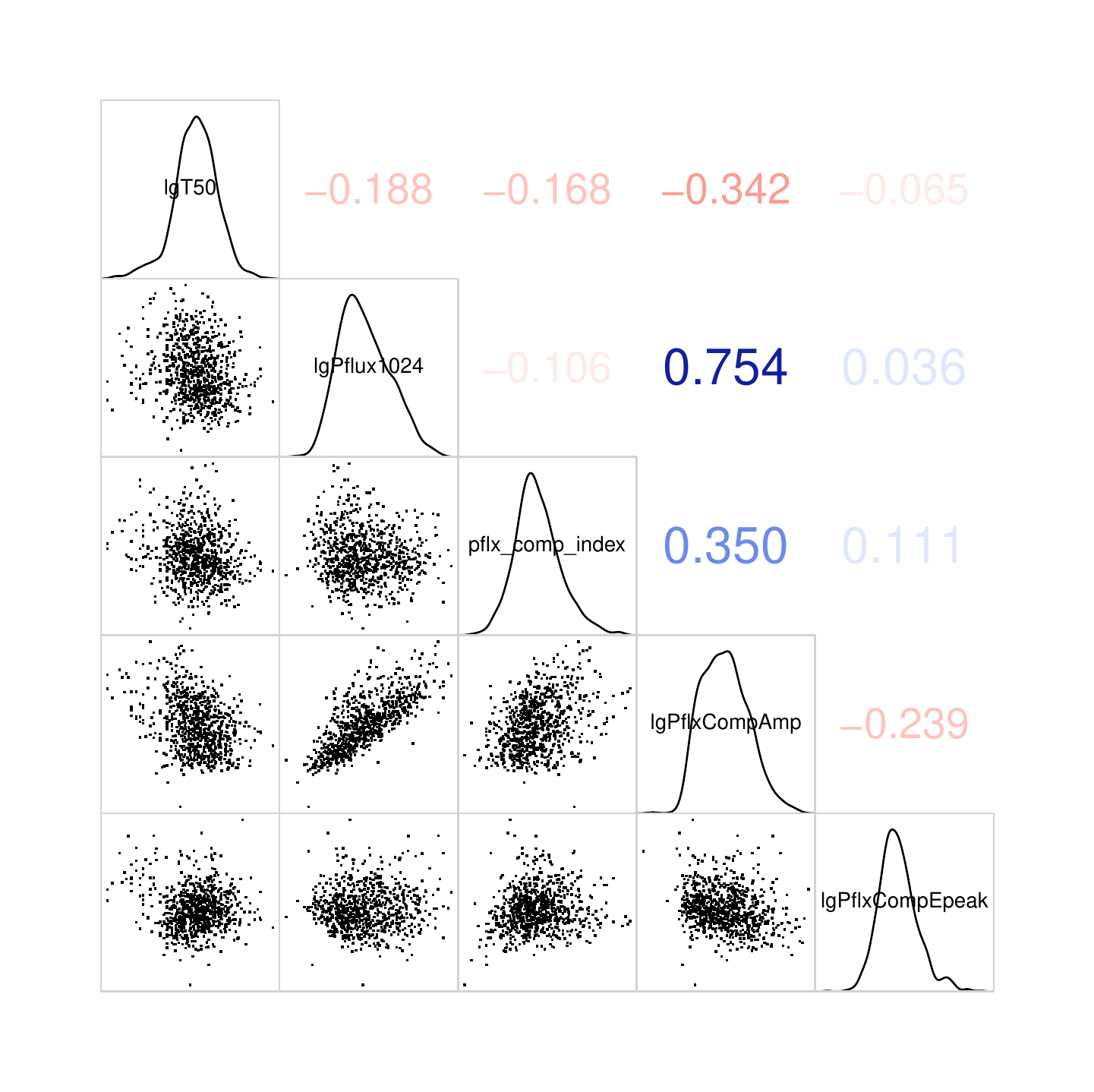} }
 \caption{\small{The 2-Dimensional distributions of lgT50, lgPflux1024, 
 pflxcompindex, lgPflxCompAmp and lgPflxCompEpeak
 and their correlation coefficients.}}
  \label{fig:n2D2}
\end{center}
\end{figure}

Table ~\ref{tab:mean} contains the mean values of each of the sixteen 
variables for all three GRB classes. Red indicates that a value
differs significantly from the other two class values in the same row.
Black numbers in a row do not differ significantly from one another. 
Blue numbers in the same row may be significantly different.
Figure ~\ref{fig:meansd} shows the mean and the standard deviation of the 
sixteen GRB classification variables for each class. 
Figure ~\ref{fig:n2D1} and Figure ~\ref{fig:n2D2} show some 
2-Dimensional distributions of our variables and also their
correlation coefficients.

\section{Robustness}
 
It is important to check the robustness of our analysis.
We do this by first excluding the variables
lgFluenceBATSE, 
lgPflux1024BATSE, 
lgPflux64BATSE and
lgPflux256BATSE. 
Second, we keep only the variables
lgT90, lgFluence, lgPflux256 and flnccompindex
which are independent of one another.

\subsection{Robustness test with twelve variables}

\begin{figure}[ht!]\begin{center}
 \resizebox{\hsize}{!}{\includegraphics[height=3.1cm,angle=0]{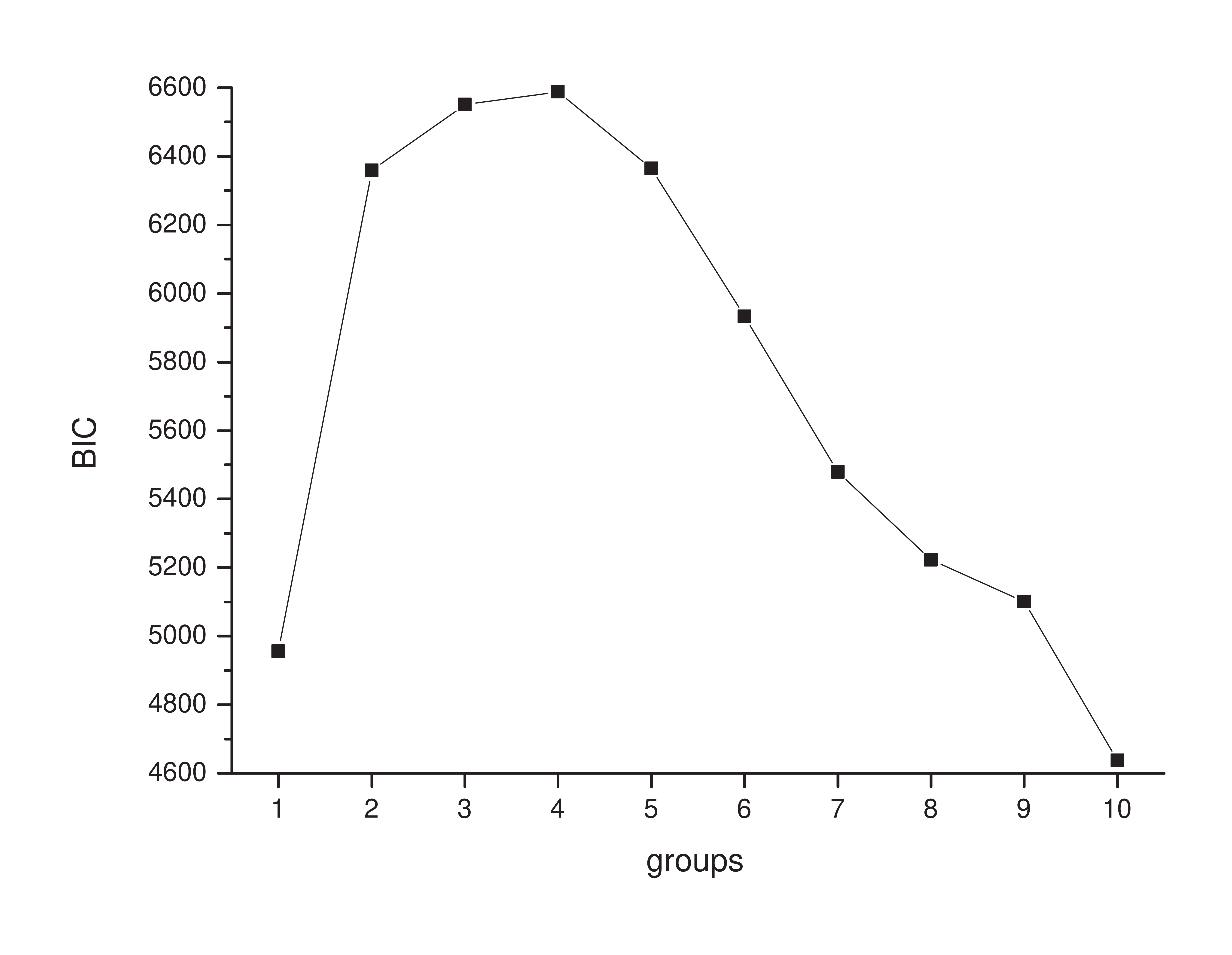} }
 \caption{\small{The Bayesian Information Criterion (BIC) suggests
that the  the Fermi GBM 12-variable data
can be describe by four classes.}}
  \label{fig:bic2}
\end{center}
\end{figure}

Excluding the BATSE-related variables
(lgFluenceBATSE, 
lgPflux1024BATSE, 
lgPflux64BATSE,
lgPflux\-256BATSE) the number of variables reduces to twelve.
We performed cluster analysis with the same 801 GRBs using the mclust() function with these twelve variables.
The BIC results can be seen in Figure~\ref{fig:bic2}. 

\begin{table}[t]\begin{center}
    \caption{Contingency table for the 16 and four variable cases with three and four groups. }
	\begin{tabular}{|l||r|r|r|r|r|}
	\hline 
\backslashbox {V16} {V12} & 1 & 2  & 3 & 4 & Total  \\ \hline \hline 
1 & 241 &  68  & 94 & 24 & 427 \\ \hline
2 & 22 &  115  & 170&  33 & 340 \\ \hline
3 & 0 &  0  &  0 & 34 & 34 \\ \hline
Total & 263 &  183  & 264 & 91 &  801 \\ \hline
	\end{tabular}
	\label{tab:contg12}
\end{center}
\end{table}

The BIC function reaches its maximum at four assumed groups. Thus, using the twelve Fermi parameters the mclust method prefers four groups rather than three.
One can compare this result with the previous result by calculating the contingency table (Table~\ref{tab:contg12}) for cases of sixteen and twelve variables with three and four GRB groups. The table shows the stability of the group members. For the first group, 56\% (241/427) remain in the same group. For the short bursts, all 34 GRBs remain in the same group (group4 in the twelve variable case).
For the second group, 84\% ((115+170)/340) remain in the new group2 and group3.

\subsection{Robustness test with four variables}

\begin{figure}[ht!]\begin{center}
 \resizebox{\hsize}{!}{\includegraphics[height=3.1cm,angle=0]{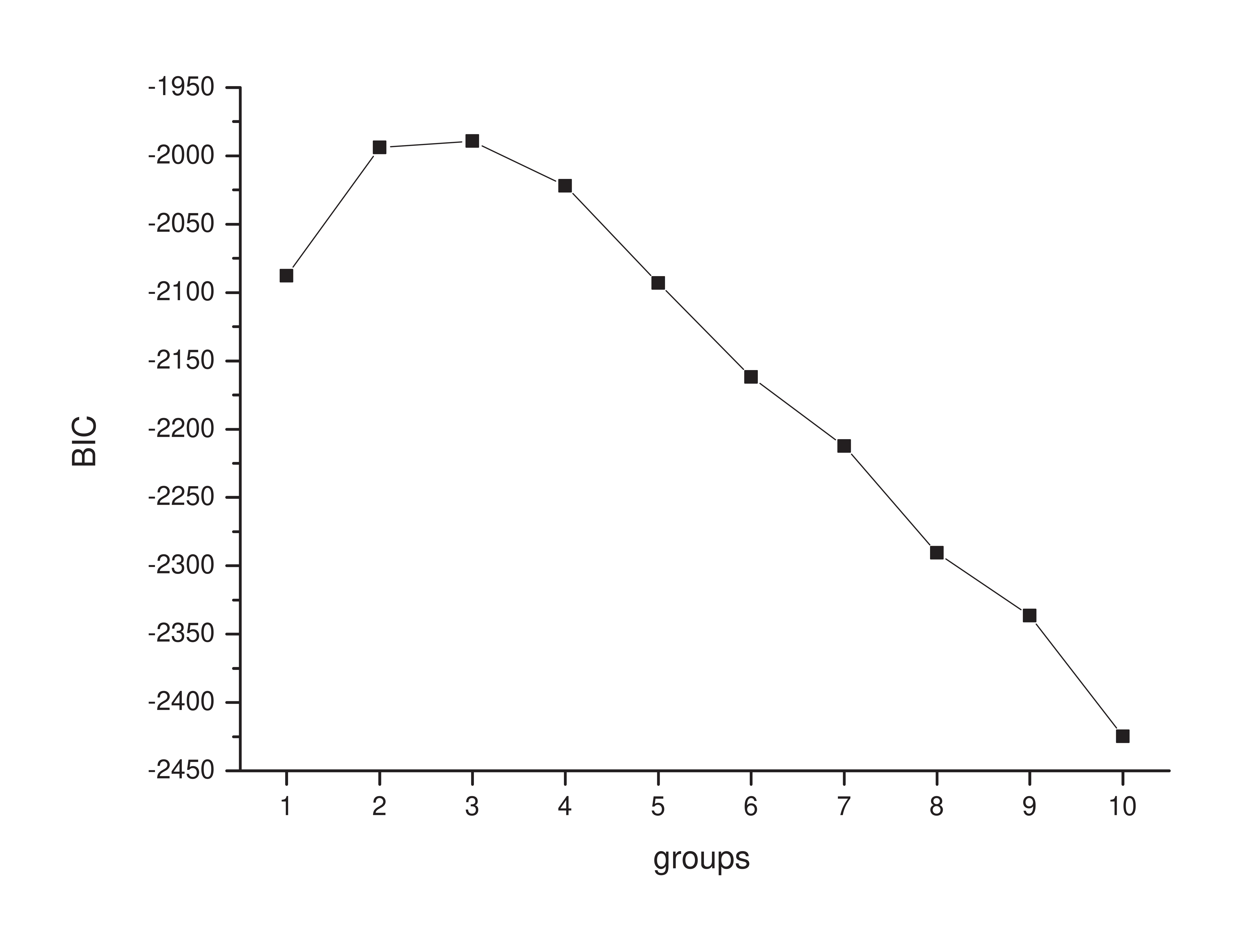} }
 
 \caption{\small{The Bayesian Information Criterion (BIC) suggests
that the  the Fermi GBM 4-variable data (lgT90, lgFluence, lgPflux256, flnccompindex)
optimally describe three GRB classes.}}
  \label{fig:bic3}
\end{center}
\end{figure}

Keeping only four variables
(lgT90, lgFluence, lgPflux256, flnccompindex)
we performed cluster analysis with the same 801 GRBs using the mclust() function. 
The BIC function reaches its maximum at three assumed groups
(see Figure~\ref{fig:bic3}).

One can compare this result with the previous result by calculating the contingency table (Table~\ref{tab:contg4}) for the sixteen and four variable cases with three assumed groups. The table shows the stabilities of the group members. For the first group, 92\% (391/427) remain in the same group. For the second and the third groups, 70\% (237/340) and 94\% (32/34) of the GRBs do not change their group membership. 

\begin{table}[t]\begin{center}
    \caption{Contingency table for the 16 and four variable cases with three groups. }
	\begin{tabular}{|l||r|r|r|r|r|}
	\hline 
\backslashbox {V16} {V4} & 1 & 2  & 3   & Total  \\ \hline \hline 
1 & 391 &  13  & 23  & 427 \\ \hline
2 & 35 &  237  & 68  & 340 \\ \hline
3 & 2 &  0  &  32   & 34 \\ \hline
Total  & 428 &  250  & 123   &  801 \\ \hline
	\end{tabular}
	\label{tab:contg4}
\end{center}
\end{table}

\section{Discussion}
Soon after the discovery of GRBs, it was suggested that they 
could be separated into two duration classes \citep{maz81,nor84}. 
This hypothesis was later supported by BATSE data 
\citep{kou93,mcb94,kos96}, and resulted in the identification of 
`Long' and `Short' GRB classes separated at around $T90=2s$.
An additional `Intermediate' class (with durations of roughly $2s  
T90 < 15s$) was subsequently found from analyses of the Third 
BATSE Catalog data \cite{muk98,hor98}; numerous 
authors \citep{hak00,bala01,rm02,hor02,hak03,bor04,hor06,chat07,zito15} 
have verified this result using BATSE GRBs. The Intermediate class 
has also been found in Beppo-SAX \citep{hor09} and Swift data 
\citep{hor08,huja09,hor10,ht16}, as well as in a preliminary 
analysis of Fermi data \cite{hor12}s. 

The Intermediate GRB class is not as clearly
delineated from the other classes as the Short 
class is; this is also true from recent analysis
of the Swift data.
\cite{zc08} analyzed 95 GRBs with measured redshifts 
and found that Swift bursts have a wider dynamic range 
in duration than pre-Swift and BATSE bursts.
\cite{kb12} subsequently analyzed the Swift BAT data and 
concluded that two classes sufficiently
describe the spectral hardness distribution, whereas three
components are needed to characterize the duration distribution.
The Intermediate class identified by \cite{kb12} has durations of around 3-20 seconds, 
which is in a good agreement with \citep{ht16} who find 
the Intermediate class durations to be in the 4 to 30 second range.

In addition to Swift, instrumental effects
might also be responsible for affecting Fermi classification results.
\cite{qin13} analyzed the data of 315 Fermi GRBs, 
studying the dependence of the duration distribution on 
energy and on various instrumental and selection effects. 
They have suggested that the true durations of a GRB could be much 
longer than what is observed.  They also suggested that the observed bimodal 
duration distribution might be due to an instrumental selection 
effect.

Analysis of data from a variety of orbital high-energy satellites
continues to find evidence for three GRB classes.
In a series of papers, Tsutsui and his coworkers have used
data from several orbital instruments, as well as
x-ray and optical afterglow data, to study GRB 
classes \citep{tsu09,tsu13,tsu14} and have found a third group with
durations of approximately five seconds. 
\cite{zito15} has analyzed the CGRO/BATSE and Swift/BAT GRB data
to find a very similar class structure to \citep{hor02}. 

Although most rigorous GRB classification studies 
find three classes in the data, there have been exceptions.
In one of his recent publications, \cite{tarno15ApSS} proposes that the division between 
short and long bursts is at 3.4 seconds rather than at two seconds. 
In \cite{tarno15AA} he analyzed the Fermi BAT duration data of 1566 GRBs. 
Although he found a third component in the distribution, the 
significance was not convincing. This may be due to methodology: he 
tested binned data with a $\chi ^2$ fit rather than using a maximum likehood 
method with unbinned data. 
Tarnopolski also suggested that the log-normal
fit may not be an adequate model for the duration distribution,
which is an interesting suggestion because
previous investigators have {\em assumed} that the
underlying distribution is lognormal.

In this manuscript we have analyzed 801 Fermi
GRBs observed by GBM using sixteen classification
variables: two durations (T90 and T50), 
three Compton spectral parameters (amplitude, peak energy, and spectral index) 
for both fluence and peak flux fits, two fluences, and 
six peak fluxes. We find that content overlap in these
variables can be reduced to a three-variable problem. 
These three main components are essentially the spectral (Comp) 
index, an amalgam of the peak fluxes, and a combination of 
fluences and durations (see Table ~\ref{tab:evect}). 
These variables can be used
to identify three clusters (classes) of GRBs. As Table ~\ref{tab:mean} and 
Figure ~\ref{fig:meansd} show, Class 3 (34 bursts) has short durations while the 
other two have long durations; Class 2 (340 bursts)
is slightly shorter than Class 1 (427 bursts). Class 2 is, however, 
brighter as measured by fluence than Class 1, which is in 
turn brighter than Class 3. The peak fluxes of Class 2 and 3 are
generally similar, and both of these are brighter than the peak 
fluxes of Class 1.  
There are no clear spectral differences between Class 1
and Class 2 bursts, although the Compton indices 
and peak energies of Class 3
are larger than those of the other two classes.

These three GRB classes are partially but not entirely recognizable 
when compared to those obtained from previous analyses.
Class 3 strongly resembles the Short class found
in the 2- and 3-class cluster analyses of other GRB experiments.
Like Short bursts, Class 3 GRBs are are less common, shorter, 
spectrally harder, and fainter than other burst classes. 
Classes 1 and 2 are not recognizable as the Intermediate and Long
burst classes found in other 3-class cluster analyses. 
Classes 1 and 2 are both equally common and of similar
spectral hardness. Class 2 is shorter yet brighter than Class 1.
In contrast, Intermediate bursts are less common, shorter, 
spectrally softer, and fainter than Long GRBs.

We have identified several reasons why our analysis
might not have delineated the Intermediate GRB
class from the Long one.
First, as indicated earlier in this section, the Intermediate class
is not as well defined as the Long class in BATSE, Swift, and Beppo-SAX
data; this may make it harder to find in the data collected by a different experiment. 
Second, Fermi GBM's smaller surface area makes it less sensitive to 
detecting faint GRBs, such as those belonging to the Intermediate class \citep{deu11}:
the relatively small number of faint but clearly identified Short GRBs 
observed by Fermi GBM supports this statement. 
Third, and most important, the spectral information presented in 
the Fermi GBM catalog is in a form that requires more interpretation 
than the hardness ratios extracted from the data of other experiments.

Spectral fit parameters are published in the Fermi GBM catalog 
rather than simpler traditional data products such as hardness ratios. 
This is because burst spectral fitting is a complex process, and the Fermi
GBM science team has chosen to publish the spectral fits 
so that burst intensity, localization, and spectral 
information have been correctly deconvolved from the detector response.
This approach serves the uninitiated user, while also
taking advantage of Fermi GBM's ability to time- and energy-tag
every photon. In contrast, other GRB experiments generally publish
bulk spectral characteristics in the form of hardness ratios,
which are simpler but not as intuitive or as potentially useful for theoretical
modeling. 
Although the Fermi GBM approach has many advantages,
it is disadvantageous when classifying GRBs
because spectral fitting requires photons to be binned in many (16) 
rather than few (4) energy channels, and there are generally too few photons
in a Fermi GBM burst for the spectrum to be fitted accurately. Since
the underlying physical GRB spectral model is not known, approximations 
are obtained in the form of the four `standard' spectral models. 
None of these models is accepted as representing the underlying burst physics, 
but each works well for burst spectra having certain characteristics 
at the cost of being less optimum for other types of spectra.
For example, the Band and SBPL models provide better spectral fits when a
significant amount of high energy flux is present, whereas
the PL and Compton models provide better fits when high
energy flux is lacking \citep{gold12,gru14}.

The choice of a `best' spectral model differs from burst to
burst, depending on factors such as the true energy distribution of photons, the
burst brightness, the burst redshift, and the detector response.
Furthermore, GRB spectra evolve, and the time interval during 
which an evolving spectrum has been observed can impact the
choice of which spectral model is optimal for fitting a burst:
rapidly evolving GRB spectra are usually spectrally harder near the time of the peak flux,
which explains why peak flux spectra might differ from fluence spectra. 
It is entirely likely and natural to think that a GRB's optimal spectral model
might change throughout its prompt emission.

However, in order to reduce measurement uncertainties in the
difficult-to-measure spectral variables, we have limited our classification database 
to those GRBs for which the Compton model provides the `best' spectral fits
from both peak fluxes and fluences. In doing so we have likely biased our sample 
to bursts having few spectral differences that might not be representative
of the larger distribution. Without having wide-ranging and well-defined burst 
spectral data, it is easy to see how our analysis might have had trouble 
clearly delineating the Intermediate class from the Long one, even as
it finds evidence supportive of three GRB classes.

\section{Summary and conclusion}
We classify GRBs using cluster analysis and 
the attributes of the published
Fermi GBM burst catalog. After initially selecting thirty-six 
potential classification variables along with 2016 
prospective bursts, 
we show that measurement uncertainties limit 
our classification sample to 810 GRBs and
only sixteen classification variables.
Principal Component Analysis reduces
the number of non-overlapping classification variables
to three, constructed primarily of peak fluxes, a spectral index,
and fluences coupled with durations.

Cluster analysis identifies three optimal GRB classes. The first
appears to be the well-known Short GRB class, while
the other two are types of Long classes. The
peak flux distributions of these two Long classes 
differ, while their hardness distributions do not
(duration and fluence  distributions are slightly different).
Neither one of these classes appears to be the 
previously-identified Intermediate class.  We attribute
this discordant result to weakly-delineated Intermediate 
class characteristics obtained from data in other GRB 
catalogs, coupled with the Fermi GBM catalog's use of 
various model-dependent spectral fitting parameters 
as opposed to standard hardness ratios.

We are currently pursuing a more detailed
analysis involving the various spectral models and
associated fitting parameters and their
effects on GRB classification. We hope to determine
(1) whether or not systematic differences in the published 
GRB spectral fitting parameters can be overcome to
yield understandable classification results, and
(2) the extent to which the Intermediate class is present
in the Fermi GBM catalog.

\begin{acknowledgements} 
This research was supported by NASA EPSCoR grant NNX13AD28A and the Hungarian TIP and TKP grants and OTKA K131653 grant. 
\end{acknowledgements}

\hyphenation{Post-Script Sprin-ger}

\end{document}